\documentclass[pra, twocolumn, amsmath, amssymb, superscriptaddress]{revtex4-2}

\let\s=\sigma

\def\bpm{\begin{pmatrix}}
\def\epm{\end{pmatrix}}
\def\be{\begin{equation}}
\def\ee{\end{equation}}
\def\bea{\begin{eqnarray}}
\def\eea{\end{eqnarray}}
\def\ba{\begin{array}}
\def\ea{\end{array}}

\def\td{\tilde}

\newcommand{\mathsym}[1]{{}}
\newcommand{\unicode}[1]{{}}

\usepackage{graphicx, braket, xcolor}
\usepackage{amsmath}
\usepackage{amssymb}
\usepackage[pdftex,colorlinks=true,citecolor=black,urlcolor=black,linkcolor=black]{hyperref}

\usepackage[inkscapelatex=false]{svg}

\usepackage[mathscr]{euscript}

\begin{document}
\title{Characterizing the transition from topology to chaos in a kicked quantum system}

\author{J. Mumford}
\affiliation{Homer L. Dodge Department of Physics and Astronomy, The University of Oklahoma, Norman, Oklahoma 73019, USA}
\affiliation{Center for Quantum Research and Technology, The University of Oklahoma, Norman, Oklahoma 73019, USA}
\author{H.-Y. Xie}
\affiliation{Homer L. Dodge Department of Physics and Astronomy, The University of Oklahoma, Norman, Oklahoma 73019, USA}
\affiliation{Center for Quantum Research and Technology, The University of Oklahoma, Norman, Oklahoma 73019, USA}
\author{R. J. Lewis-Swan}
\affiliation{Homer L. Dodge Department of Physics and Astronomy, The University of Oklahoma, Norman, Oklahoma 73019, USA}
\affiliation{Center for Quantum Research and Technology, The University of Oklahoma, Norman, Oklahoma 73019, USA}

\begin{abstract}
    This work theoretically investigates the transition from topology to chaos in a periodically driven system consisting of a quantum top coupled to a spin-1/2 particle. The system is driven by two alternating interaction kicks per period. For small kick strengths, localized topologically protected bound states exist, and as the kick strengths increase, these states proliferate.  However, at large kick strengths they gradually delocalize in stages,  eventually becoming random orthonormal vectors as chaos emerges. We identify the delocalization of the bound states as a finite size effect where their proliferation leads to their eventual overlap. This insight allows us to make analytic predictions for the onset and full emergence of chaos which are supported by numerical results of the quasi-energy level spacing ratio and Rényi entropy. A dynamical probe is also proposed to distinguish chaotic from regular behavior.
\end{abstract}

\maketitle 

\section{Introduction\label{sec:intro}}

Time-periodic driving of systems, also called Floquet engineering, has proven to be a useful tool in the study of novel phases of matter \cite{oka19}, including phenomena such as Anderson localization in time \cite{sacha15,giergiel17} and phase space crystals \cite{guo13}.
Of particular interest is the use of periodic driving of applied electromagnetic fields to study electronic materials in condensed matter physics by generating strong synthetic gauge fields and spin-orbit couplings \cite{goldman14a,goldman14b,eckardt15,bukov15} to manipulate the dispersion and geometry of Bloch-bands. This has led to the possibility of inducing topologically nontrivial phases in systems that are otherwise topologically trivial. These phases are characterized by topological invariants, such as the winding number in one-dimensional systems. Perhaps the most important feature of topological systems is the presence of localized states, also called bound states, at the interfaces between regions of different topology. These states are stable against perturbations respecting certain symmetries and protected by a large energy gap from the bulk states. The robustness of bound states to noise makes them potential candidates for encoding information in quantum computation \cite{kitaev03,alicea12,stern13}, with further applications in quantum sensing \cite{koch22}.

Concepts from topology have been extended beyond condensed matter to a wide variety of systems.  Recent experimental advancements in the control and probing of ultracold atomic gases \cite{weitenberg21} and photonic systems \cite{kim20} have allowed for the precise selection and observation of desired quantum phenomena in laboratories.  In experiments involving ultracold atomic gases, periodically driven optical lattices are able to generate large synthetic gauge fields to simulate quantum Hall physics \cite{aidelsburger13,miyake13,cooper19}.  Similarly optical waveguide and resonator arrays have been used in the study of topologically protected edge states in two and more dimensions \cite{rechtsman13,lu14,khanikaev17,ozawa19a}.  In addition, the internal states of atoms and photons are also able to mimic the spatial degrees of freedom of electrons in real materials \cite{hazzard2023}.  When periodically driven, these synthetic dimensions \cite{boada12,ozawa19b} have also been shown to host topologically nontrivial phases.  Some examples include the use of momentum \cite{meier16,xie19}, spin \cite{mancini15,stuhl15}, Fock \cite{deng22} and angular momentum \cite{cardano17} states.

Despite being driven, Floquet systems can exhibit static behavior in the form of time-averaged Hamiltonians if the driving frequency is sufficiently large, typically much larger than some microscopic energy scale \cite{abanin15,mori16}. Nevertheless, general many-body systems will eventually begin to absorb significant amounts of energy from the drive and, in the absence of any coupling to an external environment, this will lead to heating. At long times, the system thus tends toward an infinite temperature state where the topological bound states break down. This process is often accompanied by chaos, causing the once localized bound states to spread randomly throughout the Hilbert space. Efforts are typically proposed to circumvent heating \cite{bilitewski15,murakami23} as the time-averaged Hamiltonian is the primary focus of study.  However, the heating process itself has also been of interest specifically in regards to the role interactions play  \cite{lazarides14,dalessio14}.

In this paper, we investigate the transition from  topology to chaos in the context of a quantum top coupled to a spin-1/2 particle.  The system is periodically driven via two `kicks' where two different interactions between the angular momentum of the top and the spin of the particle are pulsed consecutively during each period of the drive.  It has been shown that for small kick strengths many regions of different topology, characterized by distinct winding numbers, form on the Bloch sphere of the top \cite{mumford23} and the boundaries between these regions are home to topologically protected bound states.  
In this work, we show that increasing the kick strengths results in the breakdown of these bound states, but that this breakdown 
takes place over two intermediate stages featuring progressive delocalization of the bound states, before the emergence of chaos characterized by random states.
We identify and distinguish the two intermediate stages by first the loss of the stability of the bound states to perturbations, and then the closing of the energy gap between them and the bulk states.  We use exact numerical calculations of the many-body system to 
delineate between chaos and topology using quantities including the level spacing statistics of the associated Floquet operator and the degree of localization of the bound states. Interestingly, we observe that these distinct quantities both feature qualitatively similar behavior during each step of the breakdown of the bound states.   Complementary to our numerical analysis, we also develop analytic expressions that enable us to predict the critical kick strengths at which chaos first develops and then finally overwhelms the system.  Finally, we propose a simple dynamical probe that can sensitively distinguish between the system being fully dominated by either chaos or topology.  Collectively, our findings provide insight on the transition between regimes dominated by topology and chaos in periodically driven quantum systems.


\section{Model\label{sec:model}}

\begin{figure}[t!]
    \centering
    \includegraphics[width=\columnwidth]{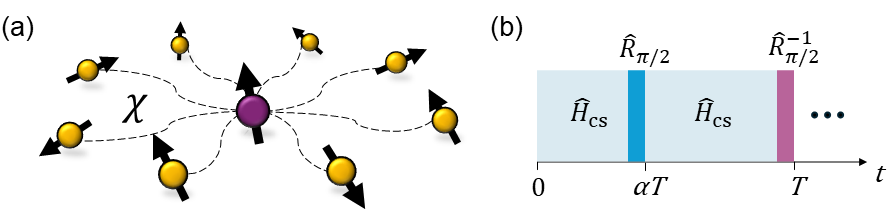}
    \caption{Central spin model and pulse sequence used to generate the Floquet operator in Eq.\ \eqref{eq:floq}.  (a) A central spin (purple) interacting equally, with strength $\chi$, with the surrounding spin-1/2 particles (yellow).  (b) The sequence relies on free evolution under the central spin Hamiltonian $\hat{H}^{(x)}_\mathrm{cs} = \chi \hat{J}_x \hat{\sigma}_x$ interrupted by two pulses at times $\alpha T$ and $T$ via the rotation operator $\hat{R}_{\pi/2} = e^{-i \left (\hat{J}_z + \hat{\sigma}_z/2\right )\pi/2}$.}
    \label{fig:floq}
\end{figure}

We consider a periodically driven bipartite system consisting of a quantum top coupled to an additional two-state degree of freedom which we will refer to as a spin-1/2 particle.  The periodicity of the driving allows the dynamics to be governed by the time-evolution operator over one period and is given by the Floquet operator,
\begin{equation}
\hat{U}_F = e^{-i \frac{\kappa_y}{j} \hat{J}_y \hat{\sigma}_y} e^{-i \frac{\kappa_x}{j} \hat{J}_x \hat{\sigma}_x}.
\label{eq:floq}
\end{equation}
Here, $\hat{J}_a$, $a=x,y,z$ is the top angular momentum operator obeying the usual commutation relation $ [\hat{J}_a, \hat{J}_b ] = i \epsilon_{abc} \hat{J}_c$ and the Pauli operators act on the spin-1/2 particle.   Each period the interactions between the top and the spin-1/2 particle are kicked twice: First with strength $\kappa_x$ between their $x$-components, then with strength $\kappa_y$ between their $y$-components.  


The Floquet operator $\hat{U}_F$ is equivalent to the unitary evolution generated by a central spin model shown in Fig.~\ref{fig:floq}(a), which describes the uniform interaction of an ensemble of qubits (yellow), equivalent to the top described above, with a central spin-1/2 particle (purple), with Hamiltonian $\hat{H}^{(a)}_{\mathrm{cs}} = \chi \hat{J}_{a} \hat{\sigma}_{a}$. Such a central spin model can be emulated in a variety of platforms, including quantum simulators composed of a Rydberg atom interacting with polar molecules \cite{dobryniecki23} or a BEC \cite{demler2019}, and is also studied in the context of NV centers \cite{chldress06} and quantum dots \cite{kessler10}. Alternating the interaction between the $x$ ($\hat{H}^{(x)}_{\mathrm{cs}}$) and $y$ ($\hat{H}^{(y)}_{\mathrm{cs}}$) frames described in Eq.~(\ref{eq:floq}) can be achieved by stroboscopically interrupting free evolution under $\hat{H}^{(a)}_{\mathrm{cs}}$ to drive the qubit and top degrees of freedom and realize rapid $\pi/2$ pulses described by the unitary operator $\hat{R}_{\pi/2} = e^{-i \left (\hat{J}_z + \hat{\sigma}_z/2\right )\pi/2}$ that switch $\hat{J}_x \leftrightarrow \hat{J}_y$ and $\hat{\sigma}_x \leftrightarrow \hat{\sigma}_y$ simultaneously [see Fig.~\ref{fig:floq}(b)].  Over one period, $T$, the desired kick strengths $\kappa_x$ and $\kappa_y$ can be achieved via free evolution under, e.g., $\hat{H}^{(x)}_\mathrm{cs}$ for a time $\alpha T$ ($\alpha \in [0,1]$) and  $\hat{H}^{(y)}_\mathrm{cs}$ for a time $(1-\alpha)T$ (with both separated by rapid $\pi/2$ pulses),
such that $\kappa_x/j = \alpha \chi T$ and $\kappa_y/j = (1-\alpha)\chi T$. Alternatively, for systems that realize a related central spin model but with XY interactions \cite{dobryniecki23}, i.e., $\hat{H} \propto \hat{J}_x \hat{\sigma}_x + \hat{J}_y \hat{\sigma}_y$, strongly driving either the top or spin-1/2 particle in a continuous fashion can also realize Eq.~(\ref{eq:floq}). Additional discussion of this is provided in Appendix \ref{sec:CS}. 

While we assume homogeneous interaction strengths between the qubits and the central spin, we will show in the next section that this condition can be relaxed without breaking the underlying symmetry of the Floquet operator that protects the topological bound states.  Inhomogeneous interactions have been studied in the context of quantum dot decoherence due to coupling with a bath of nuclear spins, where it has been shown that the decoherence time is inversely proportional to the spread of interaction strengths \cite{khaetskii02,khaetskii03,bortz07,bortz10}. Consequently, as long as the inhomogeneity remains perturbative, it will not significantly impact the dynamics of our model.

In the ring geometry displayed in Fig.\ \ref{fig:floq}(a) there are limitations on the number of qubits (yellow) that can be included before interactions between them become significant. Proposals involving polar molecules interacting with a central Rydberg atom suggest a typical range of 10–20 molecules \cite{dobryniecki23}. Accommodating more molecules is possible by placing them at a larger radius, though this reduces their interaction strength with the central atom. This reduction can be mitigated by increasing the driving period between the pulses discussed earlier.  However, the decoherence time of the central spin, along with other sources of decoherence, will impose a limit on the pulse duration.

The eigenvalues of $\hat{U}_F$ exist on the unit circle in the complex plane, so the set can be written in terms of phase factors $\{e^{-i\varepsilon_i}\}$ where $\{\varepsilon_i\}$ is the set of quasi-energies.  The term `quasi-energies' stems from the fact that $\{\varepsilon_i\}$ is obtained from the unitary operator $\hat{U}_F$, as opposed to a Hermitian operator as in the case of a typical Hamiltonian. Moreover, their values are only unique within a range of $2\pi$, for instance, $-\pi < \varepsilon_i \leq \pi$ \cite{shirley65,sambe73,grifoni98}.  This is analogous to quasi-momenta which are also $2\pi$ periodic and arise in solid state systems that are periodic in space rather than time.  In the following section we will demonstrate that the periodicity of the quasi-energy spectrum plays an important role in determining the total number of topologically protected bound states.

\section{Results\label{sec:results}}

\subsection{Topological States}

In static systems, topological phases are distinguished by the value(s) of a topological invariant such as the winding number.  In 1D topological insulators, the winding number counts the number of twists in an energy band and is therefore a global property of the system.  Changing the winding number requires the closing of the band gap and is analogous to ripping a M\"{o}bius strip to change the number of twists in it.  Through the bulk-boundary correspondence, the value of the winding number, which is a bulk quantity, is related to the number of topologically protected bound states at each boundary separating different topological regions.  These states are protected by a large energy gap between them and neighboring states in the bulk (i.e., away from the boundary).  Additional protection comes from the fact that they are eigenstates of an operator responsible for a symmetry in the system.  For instance, a Hamiltonian with chiral symmetry follows the relation $\hat{\Gamma} \hat{H} \hat{\Gamma} = - \hat{H}$ where $\hat{\Gamma}$ is the chiral symmetry operator.  This means that $\hat{\Gamma}$ takes an eigenstate of $\hat{H}$ with energy $E$ to another eigenstate with energy $-E$.  The bound states will have energy $E = 0$, so they are eigenstates of $\hat{\Gamma}$ since $\hat{\Gamma} \vert E=0 \rangle  = \pm \vert E=0 \rangle$.  

In periodically driven systems, in addition to the state with zero quasi-energy ($\varepsilon = 0$),  one finds the states at quasi-energy $\pi$ are also protected as $\hat{\Gamma} \ket{\varepsilon = \pi} = \pm \ket{\varepsilon = -\pi}$.  Here, we are interested in the symmetry of an effective Hamiltonian $\hat{H}_\mathrm{eff}$ from the Floquet operator, which can be obtained by writing $\hat{U}_F = e^{-i \hat{H}_\mathrm{eff}}$ and thus $\hat{H}_\mathrm{eff} = i \mathrm{ln}\hat{U}_F$.  However, the exact form is not obvious since one obtains an infinite number of terms if one tries to combine the two exponentials in Eq.\ \eqref{eq:floq} using the Baker-Campbell-Hausdorff formula.  Instead, it has been suggested that a Floquet system possesses chiral symmetry if there is a unitary transformation $\hat{\mathcal{U}}$ that takes $\hat{U}_F \to \hat{\tilde{U}}_F = \hat{\mathcal{U}} \hat{U}_F \hat{\mathcal{U}}^{\dagger}$ such that $\hat{\Gamma} \hat{\tilde{U}}_{F} \hat{\Gamma} = \hat{\tilde{U}}_{F}^{-1}$ \cite{asboth12,asboth13}.  In our case, there are two such 
transformations $\hat{\mathcal{U}}_1 = e^{i \frac{\kappa_y}{2j} \hat{J}_y \hat{\sigma}_y}$ and $\hat{\mathcal{U}}_2 = e^{-i \frac{\kappa_x}{2j} \hat{J}_x \hat{\sigma}_x}$ so that,
\begin{eqnarray}
\hat{\tilde{U}}_{F,1} &=& e^{-i \frac{\kappa_y}{2j} \hat{J}_y \hat{\sigma}_y} e^{-i \frac{\kappa_x}{j} \hat{J}_x \hat{\sigma}_x}e^{-i \frac{\kappa_y}{2j} \hat{J}_y \hat{\sigma}_y} , \label{eq:FC1} \\
\hat{\tilde{U}}_{F,2} &=& e^{-i \frac{\kappa_x}{2j} \hat{J}_x \hat{\sigma}_x} e^{-i \frac{\kappa_y}{j} \hat{J}_y \hat{\sigma}_y}e^{-i \frac{\kappa_x}{2j} \hat{J}_x \hat{\sigma}_x} . \label{eq:FC2}
\end{eqnarray}
It can be seen that $\hat{\Gamma} = \hat{\sigma}_z$ is an appropriate chiral symmetry operator for the symmetrized Floquet operators.  Since the chiral symmetry operator depends solely on the central spin, there is flexibility in the types of interactions between the central spin and the surrounding qubits. Notably, chiral symmetry can be preserved even without uniform interactions between the central and surrounding spins.  While any state with a quasi-energy of $0$ or $\pi$ remains protected by chiral symmetry, a large gap is still necessary for robust protection.

In addition to the chiral symmetry ($\hat{\Gamma}$), we find that the Floquet operator preserves effective parity, time-reversal ($\hat{\mathcal{T}}$), and particle-hole ($\hat{\mathcal{P}}$) symmetries. A detailed symmetry analysis in the chiral bases [Eqs.~(2) and (3)] is presented in Appendix \ref{sec:Sym}. The Floquet operator is block-diagonal in parity space. 
Each parity block of the Floquet Hamiltonian takes the symmetry operations $\hat{\mathcal{T}}^2=\hat{\mathcal{P}}^2= \hat{\Gamma}^2= 1$, which belongs to the Altland-Zirnbauer class BDI~\cite{zirnbauer96,altland97,heinzner05}. In the topological phase, the wave functions are characterized by a winding number, because the Floquet Hamiltonian effectively describes a one-dimensional tight-binding model in the Fock space spanned by the Dicke basis $\{|m,\s \rangle\}$ with $|m| \le j $ and $\s \in \{\uparrow,\downarrow\}$. 


Similar to the static case, there is a bulk-boundary correspondence in Floquet systems where the winding number counts the number of bound states at each boundary separating two or more distinct topological regions.  However, due to the fact that there are additional bound states at quasi-energy $\varepsilon = \pi$, a second winding number is needed which counts the number of those states. In our model, the physical origin of the winding numbers comes from the number of times the spin-1/2 particle completes a circuit around the equator of its Bloch sphere when the top cycles through its azimuthal angle once.  This interpretation requires the top to have a well-defined azimuthal angle to cycle through, and thus is appropriate in the limit of a large top when a mean-field or classical description is valid.
However, the winding numbers can still be calculated from the exact many-body eigenstates of $\hat{U}_F$ using the winding number operator \cite{song14,shem14} which has previously been done to show the bulk-boundary correspondence in this model \cite{mumford23}.


The locations of the topological states can be determined by taking a mean-field approximation of the top, which is in general motivated by the fact that the topological states will be well localized in the large $j$ limit such that quantum fluctuations may be neglected.  Operationally, the mean-field approximation involves replacing the operators $\hat{\pmb{J}}$ with their spin coherent state expectation values, 
\begin{equation}
\langle \theta, \phi \vert \hat{\pmb{J}} \vert \theta, \phi \rangle = j \left (\sin\theta \cos\phi, \sin\theta \sin\phi, \cos\theta \right ) , \label{eq:MF}
\end{equation}
where $\theta$ and $\phi$ are the polar and azimuthal angles of the associated Bloch sphere of the top.  Inserting Eq.~\eqref{eq:MF} into the Floquet operator removes the difficulties that come with the non-commutativity of the top operators allowing us to use the Baker-Campbell-Hausdorff formula to write $\hat{U}_F = e^{-i\hat{H}_\mathrm{eff}^\mathrm{MF}}$ where $\hat{H}_\mathrm{eff}^\mathrm{MF} = \varepsilon(\theta,\phi) \pmb{n}(\theta,\phi) \cdot \hat{\pmb{\sigma}}$ is the mean-field version of the effective Hamiltonian which can be expressed in closed form \cite{mumford23}.  The unit vector $\pmb{n}(\theta,\phi)$ points on the Bloch sphere of the spin-1/2 particle and $\varepsilon(\theta,\phi)$ is the quasi-energy

\begin{equation}
\varepsilon(\theta,\phi) = \arccos\left \{\cos \left [K_x(\theta,\phi)\right ]  \cos \left [K_y(\theta,\phi)\right ]\right \},
\label{eq:MFE}
\end{equation}
where $K_x(\theta,\phi) = \kappa_x \sin\theta\cos\phi$ and $K_y(\theta,\phi) = \kappa_y \sin\theta\sin\phi$.  From the bulk-boundary correspondence, we expect the topologically protected bound states to be located at the boundaries separating the regions of different topology.  Their locations on the Bloch sphere can be determined  by solving Eq.~\eqref{eq:MFE} when $\varepsilon(\theta,\phi) = 0,\pi$ resulting in, 


\begin{eqnarray}
z_{n_x, n_y}^{\pm} &=& \pm \sqrt{1-\pi^2 \left [\left (\frac{n_x}{\kappa_x}\right )^2 + \left (\frac{n_y}{\kappa_y} \right )^2 \right ]} \label{eq:zloc} , \\
\phi_{n_x, n_y}^{\pm \pm} &=& \pm \arccos\left [ \frac{\pm n_x/\kappa_x}{\sqrt{(n_x/\kappa_x)^2+(n_y/\kappa_y)^2 }}\right ] , \label{eq:philoc}
\end{eqnarray}
where $z = \cos\theta$ is the projection onto the $z$-axis of the Bloch sphere and $n_x, n_y \in \mathbb{Z}$.  Each pair of integers, $(n_x, n_y)$, that results in a real value of $z_{n_x,n_y}^{\pm}$ corresponds to eight bound states - one for every combination of plus and minus sign in $z_{n_x,n_y}^{\pm}$ and $\phi_{n_x,n_y}^{\pm\pm}$ (barring some special values).  Equation \eqref{eq:zloc} shows that new bound states form at the equator of the Bloch sphere of the top and as the kick strengths increase, they move toward the poles.  

\begin{figure}[t!]
    \centering
    \includegraphics[width=\columnwidth]{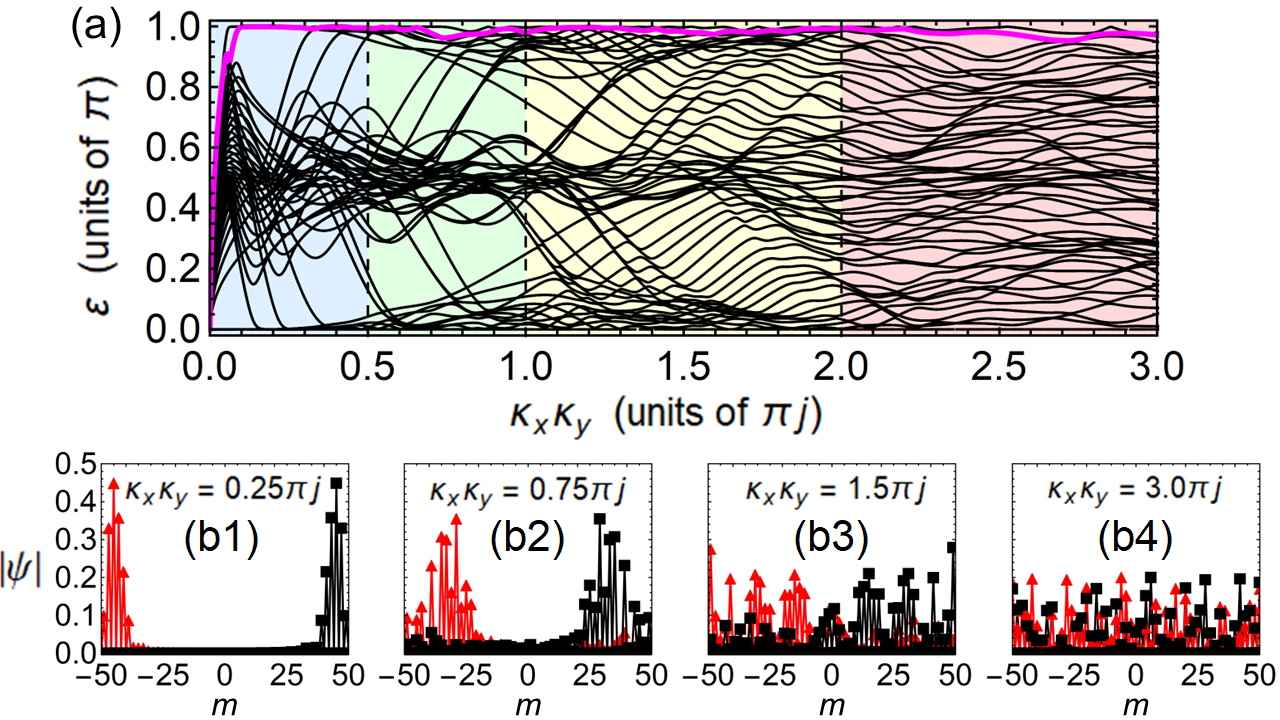}
    \caption{Quasi-energy spectrum and a topologically protected bound state for different kick strengths.  (a) The black curves are the quasi-energy spectrum and they are plotted as a function of $\kappa_x\kappa_y$.  The behavior of the spectrum is separated into four regions (indicated by the background colours): well defined bound states exist (blue); bound states lose their chiral symmetry protection (green), energy gap between the bound states and the bulk closes (yellow); bound states are destroyed and the spectrum resembles that of a chaotic system (red).  (b1)-(b4) The absolute value of the wave function of a  bound state projected onto the $\hat{J}_z$ basis with the quasi-energy of the magenta curve in (a).  The red triangles and black curves are $\vert \langle \varepsilon\vert m, \uparrow \rangle \vert$ and  $\vert \langle \varepsilon\vert m, \downarrow \rangle \vert$, respectively, which get progressively delocalized as the state goes from topological (blue) to chaotic (red).  All images are generated with $j = 50$.}
    \label{fig:PP2}
\end{figure}

\subsection{Transition From Topology to Chaos\label{sec:chaos}}

Increasing the kick strengths cannot arbitrarily increase the number of bound states in any finite system which prompts us to investigate the breakdown of this picture.  Our main result is that the bound states breakdown over several stages eventually ending up as random orthonormal vectors when chaos dominates.  While the presence of chaos at large kick strengths is not surprising, interestingly we find that the transition to chaos can be formulated entirely in terms of the breakdown of the topologically protected bound states.

Figure \ref{fig:PP2}(a) gives a qualitative overview of the different stages on the road to chaos where the quasi-energies are plotted as a function of $\kappa_x\kappa_y$.  We identify four distinct stages in the panel and characterize them in terms of the quasi-energies of the bound states.  The first stage (blue background) has well defined topological bound states at $\varepsilon = 0,\pi$, protected by both a large energy gap and chiral symmetry.  Additional bound states are generated as the kick strengths are increased, in agreement with the mean-field prediction in Eq.\ \eqref{eq:zloc} that features an increasing number of possible real solutions for larger kicks.  The new bound states form out of the bulk centered at $\varepsilon = \pi/2$ and their quasi-energies have positive or negative slopes as $\kappa_x\kappa_y$ increases depending on whether they end up with quasi-energy $\varepsilon = \pi$ or $\varepsilon = 0$, respectively.  In the second stage (green background) the quasi-energies of the majority of the bound states begin to diverge from $\varepsilon = 0, \pi$ which destroys the protection from the chiral symmetry.  This signifies the start of the breakdown of the bound states, though they are still protected by a large quasi-energy gap that persists in this stage.   In the third stage (yellow background), the gap closes, so the bound states lose all protection and begin to breakdown completely.  This is reflected in the negative and positive sloped curves going from $\varepsilon = \pi$ and $\varepsilon = 0$, respectively, to $\varepsilon = \pi/2$ signifying the return of the bound states to the bulk.  We note that the breakdown does not occur all at once, but instead gradually throughout the stage as one bound state after the other returns to the bulk.  Finally, in the fourth stage (red background), the bound and bulk states cannot be distinguished and the spectrum appears random.  Although not visible, there are avoided crossings where two quasi-energies appear to meet, which is a hallmark of the correlated energy levels found in chaotic systems.

In Fig.\ \ref{fig:PP2}(b), we plot the projection of the wave function onto the Dicke basis $\{ \vert m \rangle \}$ in each stage which corresponds to the state with the quasi-energy marked by the magenta curve at the top of panel (a).  The red triangle and black square data distinguish the projection obtained when the spin-1/2 particle is in the $\vert\uparrow\rangle$ or $\vert\downarrow\rangle$ state, respectively (see caption for details).  Starting from small kick strengths (blue background) we see that the wave function of the bound state is localized as is expected.  As the kick strengths are increased to sweep through the subsequent stages, the wave function becomes progressively more delocalized with each component of the spin-1/2 particle spreading over opposite halves of the Bloch sphere by the third stage (yellow background).  Finally, in the fourth chaotic stage (red background), the wave function is completely delocalized and resembles a random orthonormal vector.

To investigate the transition to chaos more thoroughly we analyze the quasi-energy level spacing statistics by calculating the average level spacing ratio

\begin{equation}
r = \sum_n \frac{\mathrm{min}(\delta_{n+1},\delta_n)}{\mathrm{max}(\delta_{n+1},\delta_n)},
\end{equation}
where $\delta_n = \varepsilon_{n+1} - \varepsilon_n$ is the difference between nearest quasi-energies after they are sorted in ascending order.  In integrable systems, the energy levels tend to have no correlations between them and their nearest level spacing exhibits a Poissonian distribution.  The Poissonian statistics leads to a universal value of $r$ which is $r_\mathrm{P} = 2 \mathrm{ln} 2 -1 \approx 0.386$.  In the case of chaotic systems, the energy levels are correlated and the nearest level spacing follows that of random matrices.  Here, we find that the chaotic level statistics follow random matrices from the Circular Orthogonal Ensemble (COE) such that the average level spacing ratio saturates to the universal value of $r_\mathrm{COE} = 4-2\sqrt{3} \approx 0.536$.

\begin{figure}[t!]
    \centering
    \includegraphics[width=\columnwidth]{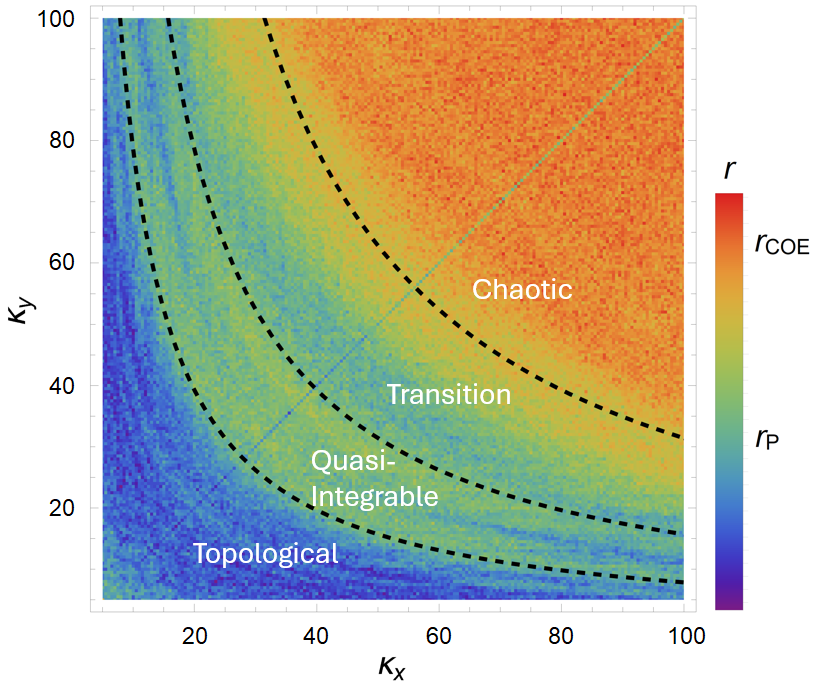}
    \caption{Density plot of the average level spacing ratio, $r$, for different kick strengths.  The plot is separated into the same four regions as in Fig.\ \ref{fig:PP2}(a) with borders at $\kappa_x\kappa_y = \pi j/2, \pi j$ and $2\pi j$ going from the bottom to the top border.  Calculation uses $j = 500$.}
    \label{fig:PD}
\end{figure}

Figure \ref{fig:PD} shows a density plot of $r$ for different kick strengths with the same distinct regions and borders as the ones shown in Fig.\ \ref{fig:PP2} (a).  The plot is symmetric about the $\kappa_y = \kappa_x$ axis because the exchange $\kappa_x \leftrightarrow \kappa_y$ can be accomplished by the unitary transformation $\hat{R}_{\pi/2}\hat{U}_F\hat{R}_{\pi/2}^\dagger$ which leaves the spectrum unchanged.  In the bottom left, for relatively small kick strengths, we have the topological region which is where well localized topologically protected bound states exist.  While remaining in this region, increasing the kick strengths also increases the number of bound states.  This is also where the mean-field predictions of the locations of the bound states in Eqns.\ \eqref{eq:zloc} and \eqref{eq:philoc} are most accurate.  $r$ falls below the Poissonian prediction due to the large degeneracy in the spectrum caused by the presence of bound states. In Appendix \ref{sec:CSB}, we demonstrate that $r$ is closer to $r_\mathrm{P}$ when the chiral symmetry is explicitly broken, which eliminates the bound states and therefore the degeneracy.  The second region is the quasi-integrable region which is where the quasi-energies of the majority of topological states are lifted from $\varepsilon = 0, \pi$, meaning these states are no longer eigenstates of $\hat{\sigma}_z$, thus they lose their protection under chiral symmetry.  The average level spacing ratio fluctuates around the integrable value of $r_\mathrm{P}$.   We have determined numerically that the transition to this region occurs when  $\left (\kappa_x \kappa_y \right )_{1} \approx  \pi (2j+1)/4$  which defines the equation of the first border.  With continued increases in kick strengths, we cross the second border and enter the transition region. Here, the average level spacing ratio gradually shifts from $r_\mathrm{P}$ to the chaotic value $r_\mathrm{COE}$ signaling a transition towards the chaotic stage.  Crossing the third border means we are in the fully chaotic region where the average level spacing ratio now stabilizes around $r_\mathrm{COE}$. Further increasing the kick strength has no consequential effect and the level spacing fluctuates around the expectation for the chaotic system. 

The positions of the second and third borders are identified by observing where the mean-field predictions first partially and then completely fail.  Our motivation for this approach comes from quantum phase transitions where the critical point separating two (or more) phases can be determined by calculating where susceptibilities predicted by the mean-field theory diverge, i.e. when the mean-field theory breaks down.  According to Eqns.\ \eqref{eq:zloc} and \eqref{eq:philoc}, we see that in the limit of $\kappa_x, \kappa_y \to \infty$, the number of topologically protected bound states also tends to infinity.  This is expected in the mean-field theory because it is exact in the thermodynamic limit ($j \to \infty$), where the number of states also tends to infinity.  However, for finite $j$, there cannot exist an infinite number of bound states, so the mean-field predictions must breakdown at some point.  To obtain an approximation for the total number of topological states predicted by the mean-field theory we assume that $\kappa_x, \kappa_y \gg 1$, so that the discrete indices in Eq.\ \eqref{eq:zloc} can be treated as continuous variables, $x = \pi n_x/\kappa_x$ and $y = \pi n_y/\kappa_y$.  Equation \eqref{eq:zloc} becomes $z(x,y) = \pm \sqrt{1-x^2-y^2}$ and the total number of topological states, $N_T$, is 

\begin{equation}
N_T \approx 8 \frac{\kappa_x\kappa_y}{\pi^2} \iiint_V dV \delta [z - z(x,y)] = 2\frac{\kappa_x\kappa_y}{\pi}
\label{eq:NT}
\end{equation}
where $dV = dxdydz$ and the volume of integration is the first octant of the unit sphere.  The factor of eight comes from the eight combinations of $\pm$ in Eqns.\ \eqref{eq:zloc} and \eqref{eq:philoc}.  We note that the factor of eight over-counts the number of bound states since for some values of $n_x$ and $n_y$ there are two instead of four solutions to Eq.\ \eqref{eq:philoc}.  An example of this is when $n_x = 0$, then we have $\phi_{0,n_y}^{\pm\pm} = \pm \arccos(0) = \pm \pi/2$.  However, the number of these types of solutions scales linearly with $\kappa_x$ and $\kappa_y$, so for $j \gg 1$ they can be neglected since we are assuming $\kappa_x, \kappa_y \gg1$ and Eq.\ \eqref{eq:NT} is proportional to the product of the two kick strengths.

We mark the second border to be where $N_T$ is equal to the total number of bulk states, thus the topologically protected bound states have essentially a second bulk themselves.  The response of the bound states to crossing this border into the transition region is to begin to breakdown into random states as seen in Fig.\ \ref{fig:PP2} (b3).  This occurs when $N_T$ is equal to half the total number of states, $2j+1$, which gives $\left (\kappa_x \kappa_y\right)_2 = \pi (2j+1)/2$ as the equation for the second border.  The third border is identified as the point where the number of bound states is equal to the total number of states, so $\left (\kappa_x \kappa_y\right )_3 = \pi (2j+1)$.  To gain some intuition on the breakdown process, one can imagine that at this point the bound states have no space between them on the Bloch sphere, so it is impossible for them to be localized.  This is reflected in Fig.\ \ref{fig:PP2} (b4) where these states have become completely delocalized resembling random orthonormal vectors which is expected in the chaotic stage.  We note that throughout the paper we have $j \gg 1$, so we approximate the border locations in the figures as $\left (\kappa_x\kappa_y\right )_1 \approx \pi j/2$, $\left (\kappa_x\kappa_y\right )_2 \approx \pi j$ and $\left (\kappa_x\kappa_y\right )_3 \approx 2\pi j$.  

\subsection{Localization of Bound States}

We can quantify the localization of the quasi-energy states by calculating the scaled R\'{e}nyi entropy
\begin{equation}
S_2(\theta,\phi) = - \frac{\ln{[\mathrm{IPR}(\theta,\phi)]}}{\ln[2(2j+1)]}.
\label{eq:RE}
\end{equation}
The function $\mathrm{IPR}(\theta,\phi) = \sum_n \vert \langle \varepsilon_n \vert \psi (\theta, \phi) \rangle \vert^4$ is the inverse participation ratio which quantifies how spread a probe state is in a given basis.  Here, we choose the probe state to be the product of a coherent spin state of the top and an equal superposition of the two states of the spin-1/2 particle $\vert \psi(\theta,\phi ) \rangle = \vert \theta, \phi \rangle \otimes \left (\vert \uparrow \rangle + \vert \downarrow \rangle \right )/\sqrt{2}$.
We take $\{\ket{\varepsilon_n}\}$ to be the basis of the Floquet operator, $\hat{U}_F$, such that when $\vert \psi(\theta,\phi) \rangle$ is an eigenstate of $\hat{U}_F$, $\mathrm{IPR} = 1$ and $S_2 = 0$, but instead when $\vert \psi(\theta,\phi)\rangle$ is spread equally throughout the entire basis, $\mathrm{IPR} = [2(2j+1)]^{-1}$ and $S_2 = 1$. In the limit of large $j$, spin coherent states have negligible quantum fluctuations (scaling as $j^{1/2}$) such that when they are centered at the same location as a topologically protected bound state (which will be one of the basis states) they will feature a strong overlap and thus $S_2$ will be small. 
However, in the chaotic stage, the bound states are destroyed and the vast majority of the quasi-energy eigenstates are delocalized over the Bloch sphere. Thus we expect that any coherent state will have a small overlap with many basis states and $S_2$ will be larger. 

\begin{figure}[t!]
    \centering
    \includegraphics[width=8cm]{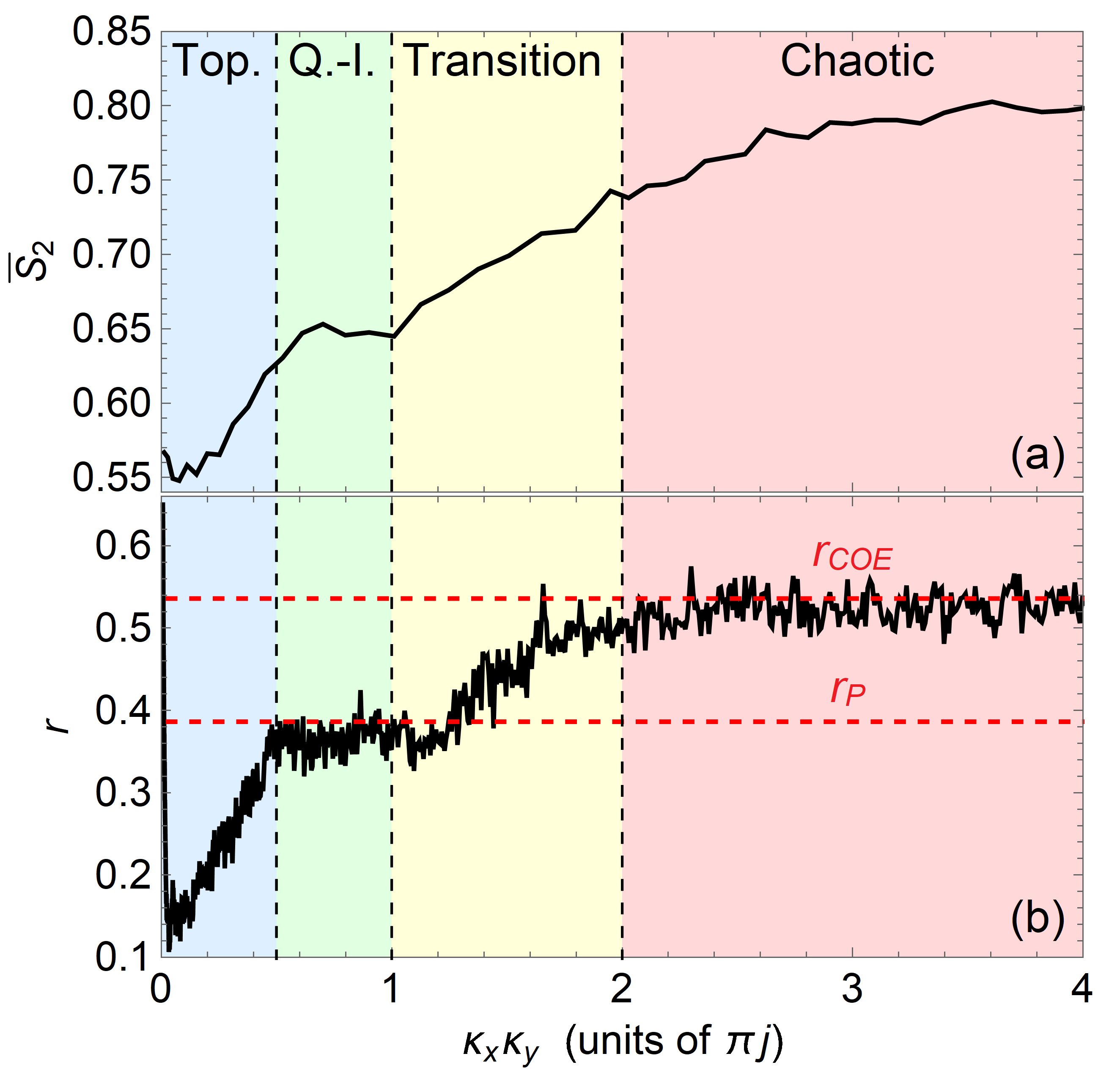}
    \caption{Average R\'{e}nyi entropy and average level spacing ratio as a function of the kick strengths.  (a) The R\'{e}nyi entropy of a coherent state averaged over the surface of the Bloch sphere of the top. (b) The average level spacing ratio of the quasi-energies.  In both panels,  $j = 250$ and the four regions (indicated by the background colors) are the same as those in Figs.\ \ref{fig:PP2} and \ref{fig:PD}.} 
    \label{fig:S2R}
\end{figure}

In Fig.\ \ref{fig:S2R} (a) we plot $S_2$ averaged over the surface of the Bloch sphere of the top, $\overline{S}_2$, as a function of $\kappa_x \kappa_y$.  We find $\overline{S}_2$ is smallest in the topological stage (blue background), consistent with the fact there are many bound states for which $S_2$ is small. As the kick strength is increased we see $\overline{S}_2$ steadily rises resulting from the bound states becoming delocalized due to the loss of the chiral symmetry protection.  This behavior leads into the quasi-integrable stage (green background) where $\overline{S}_2$ plateaus.  The apparent halt in the delocalization is attributed to the bound states still being protected by a large energy gap.  However, in the transition to chaos stage (yellow background) $\overline{S}_2$ rises again as the energy gap closes and the delocalization of the bound states (and every other state) resumes.  Finally, in the chaotic stage (red background), the states are randomly spread throughout the Hilbert space and $\overline{S}_2$ asymptotically approaches its maximum.  This value is consistent with the predictions of random matrix theory for chaotic systems, where $\mathrm{IPR}_\mathrm{COE} = \frac{3}{\mathcal{D}+2}$ \cite{sieberer19} and $\mathcal{D} = 2(2j+1)$.  Therefore, in our case $\overline{S}_2^\mathrm{COE} = \frac{\ln [(\mathcal{D}+2)/3]}{\ln [\mathcal{D}]}$ and for $j = 250$ the asymptotic value is $\overline{S}_2^\mathrm{COE} \approx 0.84$.  In addition, panel (b) displays a slice of the average level spacing ratio taken from Fig.\ \ref{fig:PD}.  Comparing this result to panel (a), we surprisingly observe that both quantities display the very similar qualitative behavior in the four stages. This indicates that either quantity can be a sensitive theoretical probe to characterize the transition from topology to chaos.

\subsection{Dynamical Probe of Transition Between Localization and Chaos}

The transition between topology and chaos can also be diagnosed from the dynamical behavior of the driven system. In particular, if a spin coherent state is initialized at the same location as a bound state, then we expect that even after many kicks the evolved state will remain close to this initial condition.
Conversely, when the system is chaotic any initial state will spread over the Bloch sphere and eventually become completely delocalized.  These two opposite behaviors can be captured in the expectation value of $\hat{J}_z$ where in the former case after $n\gg 1$ kicks we expect $\langle \hat{J}_z \rangle_n \approx \langle \hat{J}_z \rangle_0$ and in the latter case we expect $\langle \hat{J}_z \rangle_n \approx 0$ where $\langle \hat{J}_z \rangle_n = \langle \psi_0 \vert (\hat{U}_F^\dagger)^n \hat{J}_z (\hat{U}_F)^n \vert \psi_0 \rangle$. We note that for an initial state with a fixed position, the values of the kick strengths must be chosen carefully so that there exists a bound state that shares the same position.  Taking $\kappa_y$ to be a constant, we can use Eq.\ \eqref{eq:zloc} to solve for the allowed $\kappa_x$ values, $\kappa_x = \frac{n_x}{\pi}\sqrt{1-z_0^2 - \pi^2 \frac{n_y^2}{\kappa_y^2}}$ where $z_0 = \langle \hat{J}_z \rangle_0/j$.  In addition, care must be taken when choosing the initial state as, following our discussion of the stages in Fig.~\ref{fig:PD}, the bound states break down at different kick strengths. This can cause the dynamics of  $\langle \hat{J}_z \rangle_n$ to have an abrupt change in behavior that is not correlated with the predicted boundaries, but instead is away from them.  Inspecting Fig.~\ref{fig:PP2}(b1)-(b4) we note that each spin-1/2 component of a typical bound state breaks down by first spreading in the range $0 \leq m \leq \vert j\vert$, before later spreading across all values of $m$ in the chaotic stage.   Therefore, we identify that a state with $z_0= 1/2$ (i.e., $\theta_0 = \pi/3$) will be sensitive to the transition to complete chaos (red background). In the following we use 
the initial state $\vert \psi_0 \rangle  = \vert \pi/3,0\rangle \otimes \vert \uparrow \rangle$, and probe kick strengths $\kappa_x = \frac{\sqrt{3}}{2\pi}n_x$ where we have set $n_y = 0$ without loss of generality.  

\begin{figure}[t!]
    \centering
    \includegraphics[width=\columnwidth]{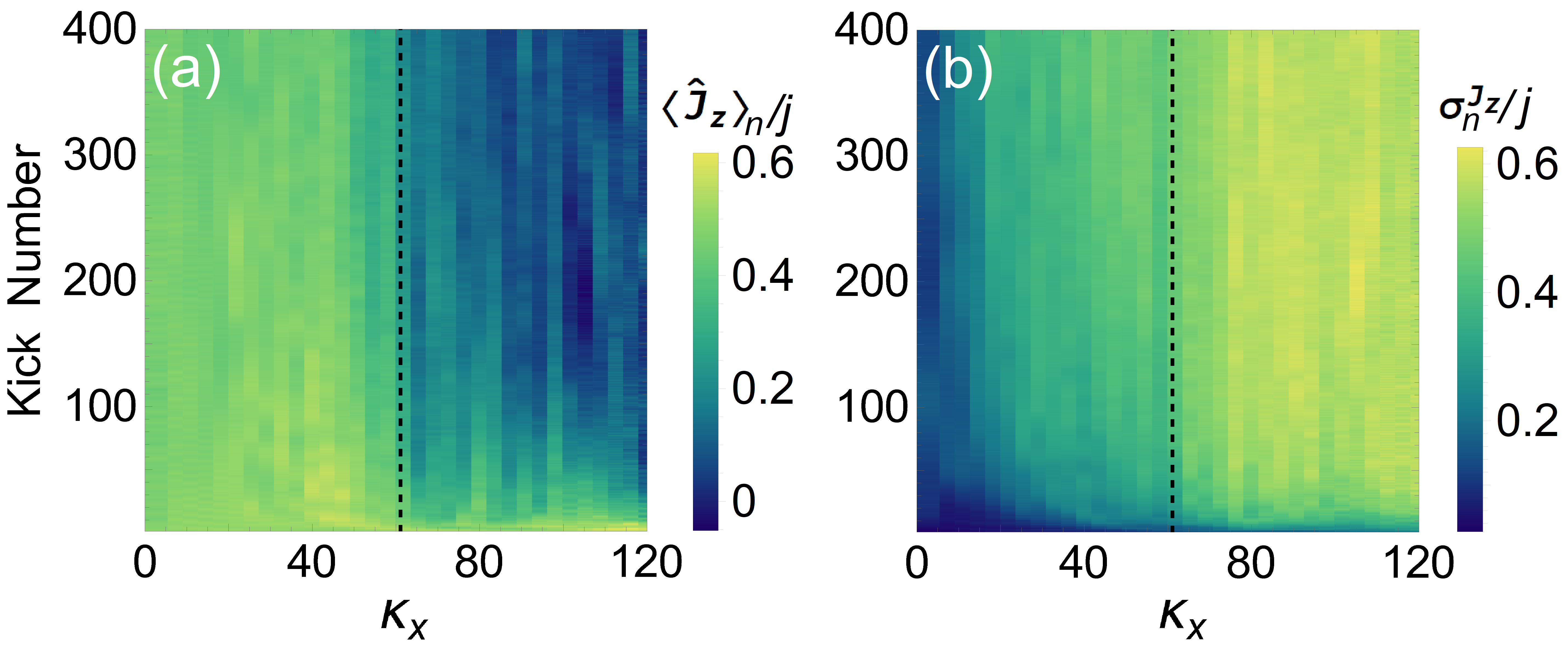}
    \caption{Density plots of the mean and standard deviation of $\hat{J}_z$ for different numbers of kicks as a function of $\kappa_x$. The initial state is $\vert \psi_0 \rangle = \vert \pi/3,0 \rangle \otimes \vert \uparrow \rangle$, $\kappa_y \approx 16.4\pi$ and $j = 500$.  The vertical dashed line marks the location of the predicted border separating the transition and chaotic stages at $\left ( \kappa_x \right)_3 =2\pi j/\kappa_y \approx 61$.}
    \label{fig:JZ}
\end{figure}

Figure \ref{fig:JZ}(a) shows a density plot of $\langle \hat{J}_z \rangle_n/j$ as a function of the number of kicks and kick strength $\kappa_x$ for $j = 500$ and $\kappa_y \approx 16.4 \pi$. To guide the eye, the vertical dashed line indicates the third border which is between the transition and chaotic stages at $\left ( \kappa_x \right )_3 = \pi(2j+1)/\kappa_y \approx 61$.  Outside of the chaotic stage, even after hundreds of kicks, $\langle \hat{J}_z \rangle _n$ stays relatively constant around its initial value of $z_0 = \langle \hat{J}_z \rangle_0/j = 1/2$, reflecting that the state remains strongly localized.  In contrast, almost immediately upon entering the chaotic stage, $\langle \hat{J}_z \rangle _n \approx 0$ after a few tens of kicks.  To confirm that the vanishing of   $\langle \hat{J}_z \rangle _n$ in the chaotic stage is due to the spreading of the wave function we also plot the standard deviation of $\hat{J}_z$, $\sigma_n^{J_z} = \sqrt{\langle \hat{J}_z^2 \rangle _n - \langle \hat{J}_z \rangle _n^2}$ for the same kick strengths in panel (b). There we see that the spreading is a smoother process, though there is a noticeable change crossing the border into the chaotic stage where  $\sigma_n^{J_z}$ remains close to the infinite temperature limit of $j/\sqrt{3}$. 

\section{Summary}

In summary, we have shown that many distinct topological regions emerge in the twice kicked top coupled to a spin-1/2 particle for weak kick strengths. Here, we have found the topologically protected bound states between each region break down in stages by becoming more delocalized as the kick strengths are increased.  For relatively large kick strengths the bound states break down completely and become random orthonormal vectors. By relating the delocalization of the bound states to the emergence of chaos, we are able to derive analytic predictions for the kick strengths when chaos first appears in the system as well as when the system becomes entirely chaotic.  These predictions are supported by numerical calculations using the average level spacing ratio of the quasi-energies as a measure of chaos and the R\'{e}nyi entropy as a measure of localization on the Bloch sphere of the top.

Unlike most condensed matter systems, which typically host only a small number of topologically protected bound states at their edges, our model allows bound states to proliferate throughout the Hilbert space. The significant fraction of bound states relative to the total number of states enabled us to investigate the emergence of chaos by focusing primarily on their localization. In contrast, a more conventional approach in Floquet systems is to identify chaos by determining when the drive frequency equals the quasi-energy bandwidth. At this point, the system often reaches an infinite-temperature steady state \cite{sen21}, which is commonly associated with chaos in many-body interacting systems \cite{deutsch91,srednicki94,srednicki99}. This raises an important question: can our method be extended to a broader class of topological systems, or is its applicability uniquely tied to the specific characteristics of our model?

\begin{acknowledgments}
This material is based upon work supported by the Air Force Office of Scientific Research under award number FA9550-24-1-0106. The computing for this project was performed at the OU Supercomputing Center for Education \& Research (OSCER) at the University of Oklahoma (OU).  The work of H.-Y.X. is supported by the Dodge Family Fellowship granted by the University of Oklahoma.
\end{acknowledgments}


\appendix

\section{Realization of $\hat{U}_F$ via a central spin model\label{sec:CS}}
In the main text we have noted that Eq.~(\ref{eq:floq}) may be realized by stroboscopically driving a top and spin-$1/2$ particle that feature a uniform interaction described by the central spin model $\hat{H}^{(a)}_{\mathrm{cs}} = \chi\hat{J}_{a} \hat{\sigma}_{a}$. Here, we provide further discussion of relevant detail for specific experimental realizations of this model, particularly focusing on generalizations of the central spin interaction.

It was recently proposed in Ref.~\cite{dobryniecki23} to use a composite system of a Rydberg atom interacting with an ensemble of polar molecules as a quantum simulator of the central spin model, although with an exchange coupling such that the simulated Hamiltonian is of the form, 
\begin{equation}
    \hat{H}_{\mathrm{int}} = \omega \hat{J}_z + \bar{\omega} \hat{\sigma}_z + \chi \left( \hat{J}^+\hat{\sigma}^- + \hat{J}^-\hat{\sigma}^+ \right) .
\end{equation}
In this case, the Floquet operator in Eq.~(\ref{eq:floq}) can instead be realized by strongly driving the spin-$1/2$ particle (or alternatively the top) with an external field whose phase is periodically jumped. In particular, we consider an applied field that drives the spin-$1/2$ particle according to the Hamiltonian, 
\begin{equation}
    \hat{H}^{\phi}_{\Omega} = \Omega_D\left[ \hat{\sigma}^+e^{i(\omega_Dt+\phi)} + \hat{\sigma}^-e^{-i(\omega_Dt+\phi)} \right] .
\end{equation}

Defining $\Delta = \bar{\omega}-\omega$ and choosing the driving field such that $\omega_D = \bar{\omega} - \Delta$, one can rewrite $\hat{H} = \hat{H}_{\mathrm{int}} + \hat{H}_{\Omega}$ in a rotating frame as, 
\begin{multline}
    \hat{H}^{\phi}_{I} = \Delta \hat{\sigma}_z + \chi \left( \hat{J}^+\hat{\sigma}^- + \hat{J}^-\hat{\sigma}^+ \right) \\ + \Omega_D\left( \hat{\sigma}^+e^{i\phi} + \hat{\sigma}^-e^{-i\phi} \right) .
\end{multline}
Moving into the interaction picture with respect to the driving terms $\hat{H}_{I,\Omega} =\Omega_D\left( \hat{\sigma}^+e^{i\phi} + \hat{\sigma}^-e^{-i\phi} \right)$ and assuming $\Omega \gg \Delta, 2j\chi$~\footnote{This condition rigorously justifies the rotating-wave approximation but when studying bound states away from the equatorial plane of the top's Bloch sphere it may be sufficient to relax it to $\Omega \gg \Delta, \sqrt{2j}\chi$.}, one can perform a rotating-wave approximation to obtain an effective Hamiltonian for the time-averaged dynamics of the system (i.e., describing dynamics coarse-grained over timescales $\delta t \gg \Omega^{-1}$) \cite{gamel10}. For example, for $\phi = 0$ one obtains, 
\begin{equation}
    \hat{H}^{\phi = 0}_{\mathrm{eff}} = e^{i\hat{H}_{I,\Omega}t} \hat{H}^{\phi = 0}_{I} e^{-i\hat{H}_{I,\Omega}t} \approx \chi \hat{J}_x \hat{\sigma}_x + \mathcal{O}(\Omega^{-2}).
\end{equation}
Similarly, for $\phi = \pi/2$ one obtains $\hat{H}^{\phi = \pi/2}_{\mathrm{eff}} = \chi \hat{J}_y\hat{\sigma}_y$. 

With the result for a general $\hat{H}^{\phi}_{\mathrm{eff}}$ in hand, we thus determine that the Floquet operator in Eq.~(\ref{eq:floq}) may be realized by: i) continuously driving the spin-$1/2$ particle with a strong external field and ii) jumping the phase once during the Floquet period between $\phi = 0$ and $\phi = \pi/2$. 

\section{Symmetries of the Floquet operator\label{sec:Sym}}
Assuming $\hat{\td{U}}_{F} \in \{\hat{\td{U}}_{F,i}\}_{i=1,2}$ in Eqs.~(2) and (3), we obtain the following symmetry properties.

\emph{Parity.} The parity operator is defined as,
\be \label{pt}
\hat{\Pi} = 
\begin{cases} 
e^{-i \pi (\hat{J}_z+\hat{\s}_z/2)}, \quad & 2j \in \mathrm{odd}, \\
i  e^{-i \pi (\hat{J}_z+\hat{\s}_z/2)}, \quad & 2j \in \mathrm{even}, 
\end{cases} \quad \hat{\Pi}^2=1.
\ee
One can show that the Floquet operator preserves parity, $\hat{\Pi} \hat{\td{U}}_{F} \hat{\Pi}^{-1} = \hat{\td{U}}_{F}$, and that it is block-diagonal in parity space.

\emph{Time-reversal symmetry.} We find two types of time-reversal operations $\hat{\mathcal{T}}_{1(2)} = \hat{T}_{1(2)} \hat{K}$ where,
\begin{align} 
&  \hat{T}_1 = 1, \quad  
\hat{\mathcal{T}}_1^2 = 1, \\
&  \hat{T}_2 = e^{-i \pi (\hat{J}_y+\hat{\s}_y/2)}, \quad 
   \hat{\mathcal{T}}_2^2 = \begin{cases}
 1, \quad & \, 2j \in   \mathrm{odd}, \\
-1, \quad & \, 2j \in  \mathrm{even}, 
\end{cases} \label{T2}
\end{align}  
and $\hat{K}$ denotes the complex conjugation operation. One can verify the time-reversal symmetry $\hat{T}_{1(2)} \hat{\td{U}}_{F}^\ast \hat{T}_{1(2)}^{-1} = \hat{\td{U}}_{F}^{-1}$. 
In addition, $\hat{\mathcal{T}}_{1}$ preserves the parity $\hat{\mathcal{T}}_{1} \hat{\Pi} \hat{\mathcal{T}}_{1}^{-1} = \hat{\Pi}$, and $\hat{\mathcal{T}}_{\mathrm{2}}$ preserves or flips the parity depending on the angular momentum of the top, $\hat{\mathcal{T}}_2 \hat{\Pi} \hat{\mathcal{T}}_{2}^{-1} = (-1)^{2j+1} \hat{\Pi}$. In parity space, $\hat{\mathcal{T}}_{1}$ is block-diagonal and $\hat{\mathcal{T}}_{2}$ is block-diagonal (block-anti-diagonal) when $2j \in$ odd (even). We note that, for $2j \in  \mathrm{even}$, Eq.~\eqref{T2} implies Kramers degeneracy. However, this Kramers degeneracy does not lead to the Circular Symplectic Ensemble level statistics in the chaotic stage, because the parity subspaces are decoupled.   

\emph{Particle-hole symmetry.} Defining the particle-hole operator $\hat{\mathcal{P}} = \hat{P}\hat{K}$, where
\be 
\hat{P} = e^{-i \frac{\pi}{2} \hat{\s}_z}, \quad \hat{\mathcal{P}}^2 = 1,
\ee
we obtain the particle-hole symmetry $\hat{P} \hat{\td{U}}_{F}^\ast \hat{P}^{-1} = \hat{\td{U}}_{F}$. This operator preserves the parity $\hat{\mathcal{P}} \hat{\Pi} \hat{\mathcal{P}}^{-1} = \hat{\Pi}$ and is block-diagonal in parity space.

\emph{Chiral symmetry.} Introducing the chiral operator
\be
\hat{\Gamma} = \hat{\s}_z, \quad \hat{\Gamma}^2=1,
\ee
we obatain the chiral symmetry $\hat{\Gamma} \hat{\td{U}}_{F} \hat{\Gamma}^{-1} = \hat{\td{U}}_{F}^{-1}$. Moreover, $\hat{\Gamma}$ preserves the parity $\hat{\Gamma} \hat{\Pi} \hat{\Gamma}^{-1} = \hat{\Pi}$, and it is block-diagonal in parity space.

Each parity block of the Floquet operator (Hamiltonian) describes an effective one-dimensional tight-binding model in the Fock space, which belongs to the
Altland-Zirnbauer class BDI with $\hat{\mathcal{T}}^2=\hat{\mathcal{P}}^2= \hat{\Gamma}^2=1$~\cite{zirnbauer96,altland97,heinzner05}. 

\section{Mean Level Spacing Ratio When Chiral Symmetry is Broken\label{sec:CSB}}

The chiral symmetry of the Floquet operator can be broken simply by adding a term proportional to $\hat{\sigma}_z$ to each kick, 

\begin{equation}
\hat{U}_F = e^{-i \left (\frac{\kappa_y}{j} \hat{J}_y \hat{\sigma}_y +\delta \hat{\sigma}_z \right )} e^{-i \left (\frac{\kappa_x}{j} \hat{J}_x \hat{\sigma}_x +\delta \hat{\sigma}_z \right )}
\label{eq:FloqC}
\end{equation}
where $\delta$ is its strength. This additional term prevents the construction of chiral-symmetry-preserving Floquet operators, such as those in Eqs. \eqref{eq:FC1} and \eqref{eq:FC2}, through unitary transformations. Breaking the chiral symmetry lifts the quasi-energies of the topologically protected bound states away from $\varepsilon =0,\pi$, ultimately leading to their destruction.  

\begin{figure}[t!]
    \centering
    \includegraphics[width=\columnwidth]{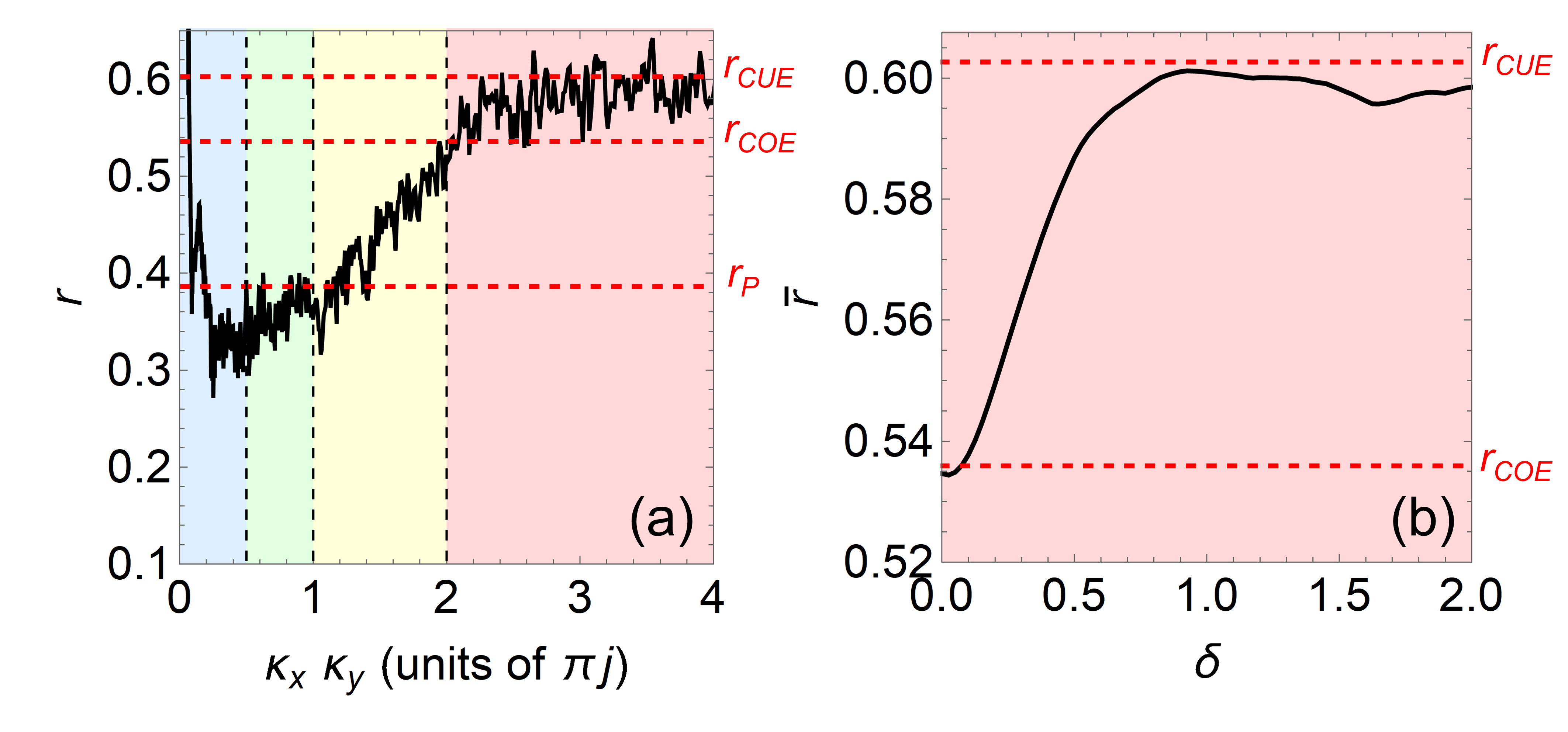}
    \caption{Mean level spacing ratio when chiral symmetry is broken. (a) Mean level spacing ratio using the quasi-energies from \eqref{eq:FloqC} as a function of $\kappa_x \kappa_y$.  The parameter values are $\delta = 1.6$ and $j = 200$.  (b)  Mean level spacing ratio averaged over the range $13 \pi j \leq \kappa_x \kappa_y \leq 20 \pi j$ (deep in the chaotic stage) as a function of $\delta$.  The background color coding and borders are the same as those used in the figures of the main text.}
    \label{fig:rrr}
\end{figure}

The effects of this symmetry breaking are evident in the mean level spacing ratio, as shown in Fig.\ \ref{fig:rrr}. Panel (a) highlights two key features. First, the initial dip and subsequent increase observed in the formerly topological stage (blue background) has nearly vanished, as the large degeneracy at quasi-energies $\varepsilon =0,\pi$ is no longer present. The level statistics shift closer to Poissonian, consistent with typical integrable systems.  Second, in the chaotic stage (red background), fluctuations no longer center around the COE prediction, but instead the Circular Unitary Ensemble (CUE) prediction of $r_\mathrm{CUE} = 2\sqrt{3}/\pi - 1/2 \approx 0.603$.  This is due to the fact that the CUE is the ensemble for matrices without any anti-unitary symmetry which is what we have for the Floquet operator when chiral symmetry is broken.  Curiously the two borders surrounding the transition stage (yellow background) still fairly accurately predict the onset, then dominance of chaos even when the system lacks substantial topology.   Panel (b) illustrates how rapidly the chiral symmetry is broken in the chaotic stage by plotting $\overline{r}$, the mean level spacing ratio averaged over the range $13 \pi j \leq \kappa_x \kappa_y \leq 20 \pi j$, as a function of $\delta$.  $\overline{r}$ starts to increase almost immediately before settling around the $r_\mathrm{CUE}$ value indicating the complete breaking of the chiral symmetry.

\bibliography{bib_TopChaos} \label{References}

\begin{thebibliography}{62}%
\makeatletter
\providecommand \@ifxundefined [1]{%
 \@ifx{#1\undefined}
}%
\providecommand \@ifnum [1]{%
 \ifnum #1\expandafter \@firstoftwo
 \else \expandafter \@secondoftwo
 \fi
}%
\providecommand \@ifx [1]{%
 \ifx #1\expandafter \@firstoftwo
 \else \expandafter \@secondoftwo
 \fi
}%
\providecommand \natexlab [1]{#1}%
\providecommand \enquote  [1]{``#1''}%
\providecommand \bibnamefont  [1]{#1}%
\providecommand \bibfnamefont [1]{#1}%
\providecommand \citenamefont [1]{#1}%
\providecommand \href@noop [0]{\@secondoftwo}%
\providecommand \href [0]{\begingroup \@sanitize@url \@href}%
\providecommand \@href[1]{\@@startlink{#1}\@@href}%
\providecommand \@@href[1]{\endgroup#1\@@endlink}%
\providecommand \@sanitize@url [0]{\catcode `\\12\catcode `\$12\catcode `\&12\catcode `\#12\catcode `\^12\catcode `\_12\catcode `\%12\relax}%
\providecommand \@@startlink[1]{}%
\providecommand \@@endlink[0]{}%
\providecommand \url  [0]{\begingroup\@sanitize@url \@url }%
\providecommand \@url [1]{\endgroup\@href {#1}{\urlprefix }}%
\providecommand \urlprefix  [0]{URL }%
\providecommand \Eprint [0]{\href }%
\providecommand \doibase [0]{https://doi.org/}%
\providecommand \selectlanguage [0]{\@gobble}%
\providecommand \bibinfo  [0]{\@secondoftwo}%
\providecommand \bibfield  [0]{\@secondoftwo}%
\providecommand \translation [1]{[#1]}%
\providecommand \BibitemOpen [0]{}%
\providecommand \bibitemStop [0]{}%
\providecommand \bibitemNoStop [0]{.\EOS\space}%
\providecommand \EOS [0]{\spacefactor3000\relax}%
\providecommand \BibitemShut  [1]{\csname bibitem#1\endcsname}%
\let\auto@bib@innerbib\@empty
\bibitem [{\citenamefont {Oka}\ and\ \citenamefont {Kitamura}(2019)}]{oka19}%
  \BibitemOpen
  \bibfield  {author} {\bibinfo {author} {\bibfnamefont {T.}~\bibnamefont {Oka}}\ and\ \bibinfo {author} {\bibfnamefont {S.}~\bibnamefont {Kitamura}},\ }\bibfield  {title} {\bibinfo {title} {{Floquet Engineering of Quantum Materials}},\ }\href {https://doi.org/10.1146/annurev-conmatphys-031218-013423} {\bibfield  {journal} {\bibinfo  {journal} {Annu. Rev. Condens. Matter Phys.}\ }\textbf {\bibinfo {volume} {10}},\ \bibinfo {pages} {387} (\bibinfo {year} {2019})}\BibitemShut {NoStop}%
\bibitem [{\citenamefont {Sacha}(2015)}]{sacha15}%
  \BibitemOpen
  \bibfield  {author} {\bibinfo {author} {\bibfnamefont {K.}~\bibnamefont {Sacha}},\ }\bibfield  {title} {\bibinfo {title} {{Anderson localization and Mott insulator phase in the time domain}},\ }\href {https://doi.org/10.1038/srep10787} {\bibfield  {journal} {\bibinfo  {journal} {Sci. Rep.}\ }\textbf {\bibinfo {volume} {5}},\ \bibinfo {pages} {10787} (\bibinfo {year} {2015})}\BibitemShut {NoStop}%
\bibitem [{\citenamefont {Giergiel}\ and\ \citenamefont {Sacha}(2017)}]{giergiel17}%
  \BibitemOpen
  \bibfield  {author} {\bibinfo {author} {\bibfnamefont {K.}~\bibnamefont {Giergiel}}\ and\ \bibinfo {author} {\bibfnamefont {K.}~\bibnamefont {Sacha}},\ }\bibfield  {title} {\bibinfo {title} {{Anderson localization of a Rydberg electron along a classical orbit}},\ }\href {https://doi.org/10.1103/PhysRevA.95.063402} {\bibfield  {journal} {\bibinfo  {journal} {Phys. Rev. A}\ }\textbf {\bibinfo {volume} {95}},\ \bibinfo {pages} {063402} (\bibinfo {year} {2017})}\BibitemShut {NoStop}%
\bibitem [{\citenamefont {Guo}\ \emph {et~al.}(2013)\citenamefont {Guo}, \citenamefont {Marthaler},\ and\ \citenamefont {Sch\"{o}n}}]{guo13}%
  \BibitemOpen
  \bibfield  {author} {\bibinfo {author} {\bibfnamefont {L.}~\bibnamefont {Guo}}, \bibinfo {author} {\bibfnamefont {M.}~\bibnamefont {Marthaler}},\ and\ \bibinfo {author} {\bibfnamefont {G.}~\bibnamefont {Sch\"{o}n}},\ }\bibfield  {title} {\bibinfo {title} {{Phase Space Crystals: A New Way to Create a Quasisenergy Band Structure}},\ }\href {https://doi.org/10.1103/PhysRevLett.111.205303} {\bibfield  {journal} {\bibinfo  {journal} {Phys. Rev. Lett.}\ }\textbf {\bibinfo {volume} {111}},\ \bibinfo {pages} {205303} (\bibinfo {year} {2013})}\BibitemShut {NoStop}%
\bibitem [{\citenamefont {Goldman}\ \emph {et~al.}(2014)\citenamefont {Goldman}, \citenamefont {Juzeli$\bar{\mathrm{u}}$nas}, \citenamefont {\"{O}hberg},\ and\ \citenamefont {Spielman}}]{goldman14a}%
  \BibitemOpen
  \bibfield  {author} {\bibinfo {author} {\bibfnamefont {N.}~\bibnamefont {Goldman}}, \bibinfo {author} {\bibfnamefont {G.}~\bibnamefont {Juzeli$\bar{\mathrm{u}}$nas}}, \bibinfo {author} {\bibfnamefont {P.}~\bibnamefont {\"{O}hberg}},\ and\ \bibinfo {author} {\bibfnamefont {I.~B.}\ \bibnamefont {Spielman}},\ }\bibfield  {title} {\bibinfo {title} {{Light-induced gauge fields for ultracold atoms}},\ }\href {https://doi.org/10.1088/0034-4885/77/12/126401} {\bibfield  {journal} {\bibinfo  {journal} {Rep. Prog. Phys.}\ }\textbf {\bibinfo {volume} {77}},\ \bibinfo {pages} {126401} (\bibinfo {year} {2014})}\BibitemShut {NoStop}%
\bibitem [{\citenamefont {Goldman}\ and\ \citenamefont {Dalibard}(2014)}]{goldman14b}%
  \BibitemOpen
  \bibfield  {author} {\bibinfo {author} {\bibfnamefont {N.}~\bibnamefont {Goldman}}\ and\ \bibinfo {author} {\bibfnamefont {J.}~\bibnamefont {Dalibard}},\ }\bibfield  {title} {\bibinfo {title} {{Periodically driven quantum systems: effective Hamiltonians and engineered gauge fields}},\ }\href {https://doi.org/10.1103/PhysRevX.4.031027} {\bibfield  {journal} {\bibinfo  {journal} {Phys. Rev. X}\ }\textbf {\bibinfo {volume} {4}},\ \bibinfo {pages} {031027} (\bibinfo {year} {2014})}\BibitemShut {NoStop}%
\bibitem [{\citenamefont {Eckardt}\ and\ \citenamefont {Anisimovas}(2015)}]{eckardt15}%
  \BibitemOpen
  \bibfield  {author} {\bibinfo {author} {\bibfnamefont {A.}~\bibnamefont {Eckardt}}\ and\ \bibinfo {author} {\bibfnamefont {E.}~\bibnamefont {Anisimovas}},\ }\bibfield  {title} {\bibinfo {title} {{High-frequency approximation for periodically driven quantum systems from a Floquet space perspective}},\ }\href {https://doi.org/10.1088/1367-2630/17/9/093039} {\bibfield  {journal} {\bibinfo  {journal} {New J. Phys.}\ }\textbf {\bibinfo {volume} {17}},\ \bibinfo {pages} {093039} (\bibinfo {year} {2015})}\BibitemShut {NoStop}%
\bibitem [{\citenamefont {Bukov}\ \emph {et~al.}(2015)\citenamefont {Bukov}, \citenamefont {D'Alessio},\ and\ \citenamefont {Polkovnikov}}]{bukov15}%
  \BibitemOpen
  \bibfield  {author} {\bibinfo {author} {\bibfnamefont {M.}~\bibnamefont {Bukov}}, \bibinfo {author} {\bibfnamefont {L.}~\bibnamefont {D'Alessio}},\ and\ \bibinfo {author} {\bibfnamefont {A.}~\bibnamefont {Polkovnikov}},\ }\bibfield  {title} {\bibinfo {title} {{Universal high-frequency behavior of periodically driven systems: from dynamical stabilization to Floquet engineering}},\ }\href {https://doi.org/10.1080/00018732.2015.1055918} {\bibfield  {journal} {\bibinfo  {journal} {Adv. Phys.}\ }\textbf {\bibinfo {volume} {64}},\ \bibinfo {pages} {129} (\bibinfo {year} {2015})}\BibitemShut {NoStop}%
\bibitem [{\citenamefont {Kitaev}(2003)}]{kitaev03}%
  \BibitemOpen
  \bibfield  {author} {\bibinfo {author} {\bibfnamefont {A.~Y.}\ \bibnamefont {Kitaev}},\ }\bibfield  {title} {\bibinfo {title} {{Fault-tolerant quantum computation by anyons}},\ }\href {https://doi.org/10.1016/S0003-4916(02)00018-0} {\bibfield  {journal} {\bibinfo  {journal} {Annals of Physics}\ }\textbf {\bibinfo {volume} {303}},\ \bibinfo {pages} {2} (\bibinfo {year} {2003})}\BibitemShut {NoStop}%
\bibitem [{\citenamefont {Alicea}(2012)}]{alicea12}%
  \BibitemOpen
  \bibfield  {author} {\bibinfo {author} {\bibfnamefont {J.}~\bibnamefont {Alicea}},\ }\bibfield  {title} {\bibinfo {title} {{New directions in the pursuit of Majorana fermions in solid state systems}},\ }\href {https://doi.org/10.1088/0034-4885/75/7/076501} {\bibfield  {journal} {\bibinfo  {journal} {Rep. Prog. Phys.}\ }\textbf {\bibinfo {volume} {75}},\ \bibinfo {pages} {076501} (\bibinfo {year} {2012})}\BibitemShut {NoStop}%
\bibitem [{\citenamefont {Stern}\ and\ \citenamefont {Lindner}(2013)}]{stern13}%
  \BibitemOpen
  \bibfield  {author} {\bibinfo {author} {\bibfnamefont {A.}~\bibnamefont {Stern}}\ and\ \bibinfo {author} {\bibfnamefont {N.~H.}\ \bibnamefont {Lindner}},\ }\bibfield  {title} {\bibinfo {title} {{Topological Quantum Computation—From Basic Concepts to First Experiments}},\ }\href {https://doi.org/10.1126/science.1231473} {\bibfield  {journal} {\bibinfo  {journal} {Science}\ }\textbf {\bibinfo {volume} {339}},\ \bibinfo {pages} {1179} (\bibinfo {year} {2013})}\BibitemShut {NoStop}%
\bibitem [{\citenamefont {Koch}\ and\ \citenamefont {Budich}(2022)}]{koch22}%
  \BibitemOpen
  \bibfield  {author} {\bibinfo {author} {\bibfnamefont {F.}~\bibnamefont {Koch}}\ and\ \bibinfo {author} {\bibfnamefont {J.~C.}\ \bibnamefont {Budich}},\ }\bibfield  {title} {\bibinfo {title} {{Quantum non-Hermitian topological sensors}},\ }\href {https://doi.org/10.1103/PhysRevResearch.4.013113} {\bibfield  {journal} {\bibinfo  {journal} {Phys. Rev. Research}\ }\textbf {\bibinfo {volume} {4}},\ \bibinfo {pages} {013113} (\bibinfo {year} {2022})}\BibitemShut {NoStop}%
\bibitem [{\citenamefont {Weitenberg}\ and\ \citenamefont {Simonet}(2021)}]{weitenberg21}%
  \BibitemOpen
  \bibfield  {author} {\bibinfo {author} {\bibfnamefont {C.}~\bibnamefont {Weitenberg}}\ and\ \bibinfo {author} {\bibfnamefont {J.}~\bibnamefont {Simonet}},\ }\bibfield  {title} {\bibinfo {title} {{Tailoring quantum gases by Floquet engineering}},\ }\href {https://doi.org/10.1038/s41567-021-01316-x} {\bibfield  {journal} {\bibinfo  {journal} {Nat. Phys.}\ }\textbf {\bibinfo {volume} {17}},\ \bibinfo {pages} {1342} (\bibinfo {year} {2021})}\BibitemShut {NoStop}%
\bibitem [{\citenamefont {Kim}\ \emph {et~al.}(2020)\citenamefont {Kim}, \citenamefont {Jacob},\ and\ \citenamefont {Rho}}]{kim20}%
  \BibitemOpen
  \bibfield  {author} {\bibinfo {author} {\bibfnamefont {M.}~\bibnamefont {Kim}}, \bibinfo {author} {\bibfnamefont {Z.}~\bibnamefont {Jacob}},\ and\ \bibinfo {author} {\bibfnamefont {J.}~\bibnamefont {Rho}},\ }\bibfield  {title} {\bibinfo {title} {{Recent advances in 2D, 3D and higher-order topological photonics}},\ }\href {https://doi.org/10.1038/s41377-020-0331-y} {\bibfield  {journal} {\bibinfo  {journal} {Light Sci. Appl.}\ }\textbf {\bibinfo {volume} {9}},\ \bibinfo {pages} {130} (\bibinfo {year} {2020})}\BibitemShut {NoStop}%
\bibitem [{\citenamefont {Aidelsburger}\ \emph {et~al.}(2013)\citenamefont {Aidelsburger}, \citenamefont {Atala}, \citenamefont {Lohse}, \citenamefont {Berreiro}, \citenamefont {Paredes},\ and\ \citenamefont {Bloch}}]{aidelsburger13}%
  \BibitemOpen
  \bibfield  {author} {\bibinfo {author} {\bibfnamefont {M.}~\bibnamefont {Aidelsburger}}, \bibinfo {author} {\bibfnamefont {M.}~\bibnamefont {Atala}}, \bibinfo {author} {\bibfnamefont {M.}~\bibnamefont {Lohse}}, \bibinfo {author} {\bibfnamefont {J.~T.}\ \bibnamefont {Berreiro}}, \bibinfo {author} {\bibfnamefont {B.}~\bibnamefont {Paredes}},\ and\ \bibinfo {author} {\bibfnamefont {I.}~\bibnamefont {Bloch}},\ }\bibfield  {title} {\bibinfo {title} {{Realization of the Hofstadter Hamiltonian with ultracold atoms in optical lattices}},\ }\href {https://doi.org/10.1103/PhysRevLett.111.185301} {\bibfield  {journal} {\bibinfo  {journal} {Phys. Rev. Lett.}\ }\textbf {\bibinfo {volume} {111}},\ \bibinfo {pages} {185301} (\bibinfo {year} {2013})}\BibitemShut {NoStop}%
\bibitem [{\citenamefont {Miyake}\ \emph {et~al.}(2013)\citenamefont {Miyake}, \citenamefont {Siviloglou}, \citenamefont {Kennedy}, \citenamefont {Burton},\ and\ \citenamefont {Ketterle}}]{miyake13}%
  \BibitemOpen
  \bibfield  {author} {\bibinfo {author} {\bibfnamefont {H.}~\bibnamefont {Miyake}}, \bibinfo {author} {\bibfnamefont {G.~A.}\ \bibnamefont {Siviloglou}}, \bibinfo {author} {\bibfnamefont {C.~J.}\ \bibnamefont {Kennedy}}, \bibinfo {author} {\bibfnamefont {W.~C.}\ \bibnamefont {Burton}},\ and\ \bibinfo {author} {\bibfnamefont {W.}~\bibnamefont {Ketterle}},\ }\bibfield  {title} {\bibinfo {title} {{Realizing the Harper Hamiltonian with laser-assisted tunneling in optical lattices}},\ }\href {https://doi.org/10.1103/PhysRevLett.111.185302} {\bibfield  {journal} {\bibinfo  {journal} {Phys. Rev. Lett.}\ }\textbf {\bibinfo {volume} {111}},\ \bibinfo {pages} {185302} (\bibinfo {year} {2013})}\BibitemShut {NoStop}%
\bibitem [{\citenamefont {Cooper}\ \emph {et~al.}(2019)\citenamefont {Cooper}, \citenamefont {Dalibard},\ and\ \citenamefont {Spielman}}]{cooper19}%
  \BibitemOpen
  \bibfield  {author} {\bibinfo {author} {\bibfnamefont {N.~R.}\ \bibnamefont {Cooper}}, \bibinfo {author} {\bibfnamefont {J.}~\bibnamefont {Dalibard}},\ and\ \bibinfo {author} {\bibfnamefont {I.~B.}\ \bibnamefont {Spielman}},\ }\bibfield  {title} {\bibinfo {title} {{Topological bands for ultracold atoms}},\ }\href {https://doi.org/10.1103/RevModPhys.91.015005} {\bibfield  {journal} {\bibinfo  {journal} {Rev. Mod. Phys.}\ }\textbf {\bibinfo {volume} {91}},\ \bibinfo {pages} {015005} (\bibinfo {year} {2019})}\BibitemShut {NoStop}%
\bibitem [{\citenamefont {Rechtsman}\ \emph {et~al.}(2013)\citenamefont {Rechtsman}, \citenamefont {Zeuner}, \citenamefont {Plotnik}, \citenamefont {Lumer}, \citenamefont {Podolsky}, \citenamefont {Dreisow}, \citenamefont {Nolte}, \citenamefont {Segev},\ and\ \citenamefont {Szameit}}]{rechtsman13}%
  \BibitemOpen
  \bibfield  {author} {\bibinfo {author} {\bibfnamefont {M.~C.}\ \bibnamefont {Rechtsman}}, \bibinfo {author} {\bibfnamefont {J.~M.}\ \bibnamefont {Zeuner}}, \bibinfo {author} {\bibfnamefont {Y.}~\bibnamefont {Plotnik}}, \bibinfo {author} {\bibfnamefont {Y.}~\bibnamefont {Lumer}}, \bibinfo {author} {\bibfnamefont {D.}~\bibnamefont {Podolsky}}, \bibinfo {author} {\bibfnamefont {F.}~\bibnamefont {Dreisow}}, \bibinfo {author} {\bibfnamefont {S.}~\bibnamefont {Nolte}}, \bibinfo {author} {\bibfnamefont {M.}~\bibnamefont {Segev}},\ and\ \bibinfo {author} {\bibfnamefont {A.}~\bibnamefont {Szameit}},\ }\bibfield  {title} {\bibinfo {title} {{Photonic Floquet topological insulators}},\ }\href {https://doi.org/10.1038/nature12066} {\bibfield  {journal} {\bibinfo  {journal} {Nature}\ }\textbf {\bibinfo {volume} {496}},\ \bibinfo {pages} {196} (\bibinfo {year} {2013})}\BibitemShut {NoStop}%
\bibitem [{\citenamefont {Ling}\ \emph {et~al.}(2014)\citenamefont {Ling}, \citenamefont {Joannopoulos},\ and\ \citenamefont {Solja\v{c}i\'{c}}}]{lu14}%
  \BibitemOpen
  \bibfield  {author} {\bibinfo {author} {\bibfnamefont {L.}~\bibnamefont {Ling}}, \bibinfo {author} {\bibfnamefont {J.~D.}\ \bibnamefont {Joannopoulos}},\ and\ \bibinfo {author} {\bibnamefont {Solja\v{c}i\'{c}}},\ }\bibfield  {title} {\bibinfo {title} {{Topological photonics}},\ }\href {https://doi.org/10.1038/nphoton.2014.248} {\bibfield  {journal} {\bibinfo  {journal} {Nat. Photon.}\ }\textbf {\bibinfo {volume} {8}},\ \bibinfo {pages} {821} (\bibinfo {year} {2014})}\BibitemShut {NoStop}%
\bibitem [{\citenamefont {Khanikaev}\ and\ \citenamefont {Shvets}(2017)}]{khanikaev17}%
  \BibitemOpen
  \bibfield  {author} {\bibinfo {author} {\bibfnamefont {A.~B.}\ \bibnamefont {Khanikaev}}\ and\ \bibinfo {author} {\bibfnamefont {G.}~\bibnamefont {Shvets}},\ }\bibfield  {title} {\bibinfo {title} {{Two-dimensional topological photonics}},\ }\href {https://doi.org/10.1038/s41566-017-0048-5} {\bibfield  {journal} {\bibinfo  {journal} {Nat. Photon.}\ }\textbf {\bibinfo {volume} {11}},\ \bibinfo {pages} {763} (\bibinfo {year} {2017})}\BibitemShut {NoStop}%
\bibitem [{\citenamefont {Ozawa}\ \emph {et~al.}(2019)\citenamefont {Ozawa}, \citenamefont {Price}, \citenamefont {Amo}, \citenamefont {Goldman}, \citenamefont {Hafezi}, \citenamefont {Lu}, \citenamefont {Rechtsman}, \citenamefont {Schuster}, \citenamefont {Simon}, \citenamefont {Zilbergberg},\ and\ \citenamefont {Carusotto}}]{ozawa19a}%
  \BibitemOpen
  \bibfield  {author} {\bibinfo {author} {\bibfnamefont {T.}~\bibnamefont {Ozawa}}, \bibinfo {author} {\bibfnamefont {H.~M.}\ \bibnamefont {Price}}, \bibinfo {author} {\bibfnamefont {A.}~\bibnamefont {Amo}}, \bibinfo {author} {\bibfnamefont {N.}~\bibnamefont {Goldman}}, \bibinfo {author} {\bibfnamefont {M.}~\bibnamefont {Hafezi}}, \bibinfo {author} {\bibfnamefont {L.}~\bibnamefont {Lu}}, \bibinfo {author} {\bibfnamefont {M.~C.}\ \bibnamefont {Rechtsman}}, \bibinfo {author} {\bibfnamefont {D.}~\bibnamefont {Schuster}}, \bibinfo {author} {\bibfnamefont {J.}~\bibnamefont {Simon}}, \bibinfo {author} {\bibfnamefont {O.}~\bibnamefont {Zilbergberg}},\ and\ \bibinfo {author} {\bibfnamefont {I.}~\bibnamefont {Carusotto}},\ }\bibfield  {title} {\bibinfo {title} {{Topological photonics}},\ }\href {https://doi.org/10.1103/RevModPhys.91.015006} {\bibfield  {journal} {\bibinfo  {journal} {Rev. Mod. Phys.}\ }\textbf {\bibinfo {volume} {91}},\ \bibinfo {pages} {015006} (\bibinfo {year} {2019})}\BibitemShut {NoStop}%
\bibitem [{\citenamefont {Hazzard}\ and\ \citenamefont {Gadway}(2023)}]{hazzard2023}%
  \BibitemOpen
  \bibfield  {author} {\bibinfo {author} {\bibfnamefont {K.~R.~A.}\ \bibnamefont {Hazzard}}\ and\ \bibinfo {author} {\bibfnamefont {B.}~\bibnamefont {Gadway}},\ }\bibfield  {title} {\bibinfo {title} {{Synthetic dimensions}},\ }\href {https://doi.org/10.1063/PT.3.5225} {\bibfield  {journal} {\bibinfo  {journal} {Physics Today}\ }\textbf {\bibinfo {volume} {76}},\ \bibinfo {pages} {62} (\bibinfo {year} {2023})}\BibitemShut {NoStop}%
\bibitem [{\citenamefont {Boada}\ \emph {et~al.}(2012)\citenamefont {Boada}, \citenamefont {Celi}, \citenamefont {Latorre},\ and\ \citenamefont {Lewenstein}}]{boada12}%
  \BibitemOpen
  \bibfield  {author} {\bibinfo {author} {\bibfnamefont {O.}~\bibnamefont {Boada}}, \bibinfo {author} {\bibfnamefont {A.}~\bibnamefont {Celi}}, \bibinfo {author} {\bibfnamefont {J.~I.}\ \bibnamefont {Latorre}},\ and\ \bibinfo {author} {\bibfnamefont {M.}~\bibnamefont {Lewenstein}},\ }\bibfield  {title} {\bibinfo {title} {{Quantum simulation of an extra dimension}},\ }\href {https://doi.org/10.1103/PhysRevLett.108.133001} {\bibfield  {journal} {\bibinfo  {journal} {Phys. Rev. Lett.}\ }\textbf {\bibinfo {volume} {108}},\ \bibinfo {pages} {133001} (\bibinfo {year} {2012})}\BibitemShut {NoStop}%
\bibitem [{\citenamefont {Ozawa}\ and\ \citenamefont {Price}(2019)}]{ozawa19b}%
  \BibitemOpen
  \bibfield  {author} {\bibinfo {author} {\bibfnamefont {T.}~\bibnamefont {Ozawa}}\ and\ \bibinfo {author} {\bibfnamefont {H.~M.}\ \bibnamefont {Price}},\ }\bibfield  {title} {\bibinfo {title} {{Topological quantum matter in synthetic dimensions}},\ }\href {https://doi.org/10.1038/s42254-019-0045-3} {\bibfield  {journal} {\bibinfo  {journal} {Nat. Rev. Phys.}\ }\textbf {\bibinfo {volume} {1}},\ \bibinfo {pages} {349–} (\bibinfo {year} {2019})}\BibitemShut {NoStop}%
\bibitem [{\citenamefont {Meier}\ \emph {et~al.}(2016)\citenamefont {Meier}, \citenamefont {An},\ and\ \citenamefont {Gadway}}]{meier16}%
  \BibitemOpen
  \bibfield  {author} {\bibinfo {author} {\bibfnamefont {E.~J.}\ \bibnamefont {Meier}}, \bibinfo {author} {\bibfnamefont {F.~A.}\ \bibnamefont {An}},\ and\ \bibinfo {author} {\bibfnamefont {B.}~\bibnamefont {Gadway}},\ }\bibfield  {title} {\bibinfo {title} {{Observation of the topological soliton state in the Su-Schrieffer-Heeger model}},\ }\href {https://doi.org/10.1038/ncomms13986} {\bibfield  {journal} {\bibinfo  {journal} {Nat. Comm.}\ }\textbf {\bibinfo {volume} {7}},\ \bibinfo {pages} {13986} (\bibinfo {year} {2016})}\BibitemShut {NoStop}%
\bibitem [{\citenamefont {Xie}\ \emph {et~al.}(2019)\citenamefont {Xie}, \citenamefont {Gou}, \citenamefont {Xiao}, \citenamefont {Gadway},\ and\ \citenamefont {Yan}}]{xie19}%
  \BibitemOpen
  \bibfield  {author} {\bibinfo {author} {\bibfnamefont {D.}~\bibnamefont {Xie}}, \bibinfo {author} {\bibfnamefont {W.}~\bibnamefont {Gou}}, \bibinfo {author} {\bibfnamefont {T.}~\bibnamefont {Xiao}}, \bibinfo {author} {\bibfnamefont {B.}~\bibnamefont {Gadway}},\ and\ \bibinfo {author} {\bibfnamefont {B.}~\bibnamefont {Yan}},\ }\bibfield  {title} {\bibinfo {title} {{Topological characterizations of an extended Su-Schrieffer-Heeger model}},\ }\href@noop {} {\bibfield  {journal} {\bibinfo  {journal} {npj Quantum Inf.}\ }\textbf {\bibinfo {volume} {5}},\ \bibinfo {pages} {55} (\bibinfo {year} {2019})}\BibitemShut {NoStop}%
\bibitem [{\citenamefont {Mancini}\ \emph {et~al.}(2015)\citenamefont {Mancini}, \citenamefont {Pagano}, \citenamefont {Cappellini}, \citenamefont {Livi}, \citenamefont {Rider}, \citenamefont {Catani}, \citenamefont {Sias}, \citenamefont {Zoller}, \citenamefont {Inguscio}, \citenamefont {Dalmonte},\ and\ \citenamefont {Fallani}}]{mancini15}%
  \BibitemOpen
  \bibfield  {author} {\bibinfo {author} {\bibfnamefont {M.}~\bibnamefont {Mancini}}, \bibinfo {author} {\bibfnamefont {G.}~\bibnamefont {Pagano}}, \bibinfo {author} {\bibfnamefont {G.}~\bibnamefont {Cappellini}}, \bibinfo {author} {\bibfnamefont {L.}~\bibnamefont {Livi}}, \bibinfo {author} {\bibfnamefont {M.}~\bibnamefont {Rider}}, \bibinfo {author} {\bibfnamefont {J.}~\bibnamefont {Catani}}, \bibinfo {author} {\bibfnamefont {C.}~\bibnamefont {Sias}}, \bibinfo {author} {\bibfnamefont {P.}~\bibnamefont {Zoller}}, \bibinfo {author} {\bibfnamefont {M.}~\bibnamefont {Inguscio}}, \bibinfo {author} {\bibfnamefont {M.}~\bibnamefont {Dalmonte}},\ and\ \bibinfo {author} {\bibfnamefont {L.}~\bibnamefont {Fallani}},\ }\bibfield  {title} {\bibinfo {title} {{Observation of chiral edge states with neutral fermions in synthetic Hall ribbons}},\ }\href {https://doi.org/10.1126/science.aaa8736} {\bibfield  {journal} {\bibinfo  {journal} {Science}\ }\textbf {\bibinfo {volume} {349}},\ \bibinfo {pages} {1510} (\bibinfo {year}
  {2015})}\BibitemShut {NoStop}%
\bibitem [{\citenamefont {Stuhl}\ \emph {et~al.}(2015)\citenamefont {Stuhl}, \citenamefont {Lu}, \citenamefont {Aycock}, \citenamefont {Genkina},\ and\ \citenamefont {Spielman}}]{stuhl15}%
  \BibitemOpen
  \bibfield  {author} {\bibinfo {author} {\bibfnamefont {B.~K.}\ \bibnamefont {Stuhl}}, \bibinfo {author} {\bibfnamefont {H.-I.}\ \bibnamefont {Lu}}, \bibinfo {author} {\bibfnamefont {L.~M.}\ \bibnamefont {Aycock}}, \bibinfo {author} {\bibfnamefont {D.}~\bibnamefont {Genkina}},\ and\ \bibinfo {author} {\bibfnamefont {I.~B.}\ \bibnamefont {Spielman}},\ }\bibfield  {title} {\bibinfo {title} {{Visualizing edge states with an atomic Bose gas in the quantum Hall regime}},\ }\href {https://doi.org/10.1126/science.aaa8515} {\bibfield  {journal} {\bibinfo  {journal} {Science}\ }\textbf {\bibinfo {volume} {349}},\ \bibinfo {pages} {1514} (\bibinfo {year} {2015})}\BibitemShut {NoStop}%
\bibitem [{\citenamefont {Deng}\ \emph {et~al.}(2022)\citenamefont {Deng}, \citenamefont {Dong}, \citenamefont {Zhang}, \citenamefont {Wu}, \citenamefont {Yuan}, \citenamefont {Zhu}, \citenamefont {Jin}, \citenamefont {Li}, \citenamefont {Wang}, \citenamefont {Cai}, \citenamefont {Song}, \citenamefont {Wang}, \citenamefont {You},\ and\ \citenamefont {Wang}}]{deng22}%
  \BibitemOpen
  \bibfield  {author} {\bibinfo {author} {\bibfnamefont {J.}~\bibnamefont {Deng}}, \bibinfo {author} {\bibfnamefont {H.}~\bibnamefont {Dong}}, \bibinfo {author} {\bibfnamefont {C.}~\bibnamefont {Zhang}}, \bibinfo {author} {\bibfnamefont {Y.}~\bibnamefont {Wu}}, \bibinfo {author} {\bibfnamefont {J.}~\bibnamefont {Yuan}}, \bibinfo {author} {\bibfnamefont {X.}~\bibnamefont {Zhu}}, \bibinfo {author} {\bibfnamefont {F.}~\bibnamefont {Jin}}, \bibinfo {author} {\bibfnamefont {H.}~\bibnamefont {Li}}, \bibinfo {author} {\bibfnamefont {Z.}~\bibnamefont {Wang}}, \bibinfo {author} {\bibfnamefont {H.}~\bibnamefont {Cai}}, \bibinfo {author} {\bibfnamefont {C.}~\bibnamefont {Song}}, \bibinfo {author} {\bibfnamefont {H.}~\bibnamefont {Wang}}, \bibinfo {author} {\bibfnamefont {J.~Q.}\ \bibnamefont {You}},\ and\ \bibinfo {author} {\bibfnamefont {D.-W.}\ \bibnamefont {Wang}},\ }\bibfield  {title} {\bibinfo {title} {{Observing the quantum topology of light}},\ }\href {https://doi.org/10.1126/science.ade6219} {\bibfield
  {journal} {\bibinfo  {journal} {Science}\ }\textbf {\bibinfo {volume} {378}},\ \bibinfo {pages} {966} (\bibinfo {year} {2022})}\BibitemShut {NoStop}%
\bibitem [{\citenamefont {Cardano}\ \emph {et~al.}(2017)\citenamefont {Cardano}, \citenamefont {D'Errico}, \citenamefont {Dauphin}, \citenamefont {Maffei}, \citenamefont {Piccirillo}, \citenamefont {de~Lisio}, \citenamefont {De~Filippis}, \citenamefont {Cataudella}, \citenamefont {Santamato}, \citenamefont {Marrucci}, \citenamefont {Lewenstein},\ and\ \citenamefont {Massignan}}]{cardano17}%
  \BibitemOpen
  \bibfield  {author} {\bibinfo {author} {\bibfnamefont {F.}~\bibnamefont {Cardano}}, \bibinfo {author} {\bibfnamefont {A.}~\bibnamefont {D'Errico}}, \bibinfo {author} {\bibfnamefont {A.}~\bibnamefont {Dauphin}}, \bibinfo {author} {\bibfnamefont {M.}~\bibnamefont {Maffei}}, \bibinfo {author} {\bibfnamefont {B.}~\bibnamefont {Piccirillo}}, \bibinfo {author} {\bibfnamefont {C.}~\bibnamefont {de~Lisio}}, \bibinfo {author} {\bibfnamefont {G.}~\bibnamefont {De~Filippis}}, \bibinfo {author} {\bibfnamefont {V.}~\bibnamefont {Cataudella}}, \bibinfo {author} {\bibfnamefont {E.}~\bibnamefont {Santamato}}, \bibinfo {author} {\bibfnamefont {L.}~\bibnamefont {Marrucci}}, \bibinfo {author} {\bibfnamefont {M.}~\bibnamefont {Lewenstein}},\ and\ \bibinfo {author} {\bibfnamefont {P.}~\bibnamefont {Massignan}},\ }\bibfield  {title} {\bibinfo {title} {{Detection of Zak phases and topological invariants in a chiral quantum walk of twisted photons}},\ }\href {https://doi.org/10.1038/ncomms15516} {\bibfield  {journal} {\bibinfo
  {journal} {Nat. Comm.}\ }\textbf {\bibinfo {volume} {8}},\ \bibinfo {pages} {15516} (\bibinfo {year} {2017})}\BibitemShut {NoStop}%
\bibitem [{\citenamefont {Abanin}\ \emph {et~al.}(2015)\citenamefont {Abanin}, \citenamefont {De~Roeck},\ and\ \citenamefont {Huveneers}}]{abanin15}%
  \BibitemOpen
  \bibfield  {author} {\bibinfo {author} {\bibfnamefont {D.~A.}\ \bibnamefont {Abanin}}, \bibinfo {author} {\bibfnamefont {W.}~\bibnamefont {De~Roeck}},\ and\ \bibinfo {author} {\bibfnamefont {F.~m.~c.}\ \bibnamefont {Huveneers}},\ }\bibfield  {title} {\bibinfo {title} {{Exponentially Slow Heating in Periodically Driven Many-Body Systems}},\ }\href {https://doi.org/10.1103/PhysRevLett.115.256803} {\bibfield  {journal} {\bibinfo  {journal} {Phys. Rev. Lett.}\ }\textbf {\bibinfo {volume} {115}},\ \bibinfo {pages} {256803} (\bibinfo {year} {2015})}\BibitemShut {NoStop}%
\bibitem [{\citenamefont {Mori}\ \emph {et~al.}(2016)\citenamefont {Mori}, \citenamefont {Kuwahara},\ and\ \citenamefont {Saito}}]{mori16}%
  \BibitemOpen
  \bibfield  {author} {\bibinfo {author} {\bibfnamefont {T.}~\bibnamefont {Mori}}, \bibinfo {author} {\bibfnamefont {T.}~\bibnamefont {Kuwahara}},\ and\ \bibinfo {author} {\bibfnamefont {K.}~\bibnamefont {Saito}},\ }\bibfield  {title} {\bibinfo {title} {{Rigorous Bound on Energy Absorption and Generic Relaxation in Periodically Driven Quantum Systems}},\ }\href {https://doi.org/10.1103/PhysRevLett.116.120401} {\bibfield  {journal} {\bibinfo  {journal} {Phys. Rev. Lett.}\ }\textbf {\bibinfo {volume} {116}},\ \bibinfo {pages} {120401} (\bibinfo {year} {2016})}\BibitemShut {NoStop}%
\bibitem [{\citenamefont {Bilitewski}\ and\ \citenamefont {Cooper}(2015)}]{bilitewski15}%
  \BibitemOpen
  \bibfield  {author} {\bibinfo {author} {\bibfnamefont {T.}~\bibnamefont {Bilitewski}}\ and\ \bibinfo {author} {\bibfnamefont {N.~R.}\ \bibnamefont {Cooper}},\ }\bibfield  {title} {\bibinfo {title} {{Population dynamics in a Floquet realization of the Harper-Hofstadter Hamiltonian}},\ }\href {https://doi.org/10.1103/PhysRevA.91.063611} {\bibfield  {journal} {\bibinfo  {journal} {Phys. Rev. A}\ }\textbf {\bibinfo {volume} {91}},\ \bibinfo {pages} {063611} (\bibinfo {year} {2015})}\BibitemShut {NoStop}%
\bibitem [{\citenamefont {Murakami}\ \emph {et~al.}(2023)\citenamefont {Murakami}, \citenamefont {Sch\"{u}ler}, \citenamefont {Arita},\ and\ \citenamefont {Werner}}]{murakami23}%
  \BibitemOpen
  \bibfield  {author} {\bibinfo {author} {\bibfnamefont {Y.}~\bibnamefont {Murakami}}, \bibinfo {author} {\bibfnamefont {M.}~\bibnamefont {Sch\"{u}ler}}, \bibinfo {author} {\bibfnamefont {R.}~\bibnamefont {Arita}},\ and\ \bibinfo {author} {\bibfnamefont {P.}~\bibnamefont {Werner}},\ }\bibfield  {title} {\bibinfo {title} {{Suppression of heating by multicolor driving protocols in Floquet-engineered strongly correlated systems}},\ }\href {https://doi.org/10.1103/PhysRevB.108.035151} {\bibfield  {journal} {\bibinfo  {journal} {Phys. Rev. B}\ }\textbf {\bibinfo {volume} {108}},\ \bibinfo {pages} {035151} (\bibinfo {year} {2023})}\BibitemShut {NoStop}%
\bibitem [{\citenamefont {Lazarides}\ \emph {et~al.}(2014)\citenamefont {Lazarides}, \citenamefont {Das},\ and\ \citenamefont {Moessner}}]{lazarides14}%
  \BibitemOpen
  \bibfield  {author} {\bibinfo {author} {\bibfnamefont {A.}~\bibnamefont {Lazarides}}, \bibinfo {author} {\bibfnamefont {A.}~\bibnamefont {Das}},\ and\ \bibinfo {author} {\bibfnamefont {R.}~\bibnamefont {Moessner}},\ }\bibfield  {title} {\bibinfo {title} {{Equilibrium states of generic quantum systems subject to periodic driving}},\ }\href {https://doi.org/10.1103/PhysRevE.90.012110} {\bibfield  {journal} {\bibinfo  {journal} {Phys. Rev. B}\ }\textbf {\bibinfo {volume} {90}},\ \bibinfo {pages} {012110} (\bibinfo {year} {2014})}\BibitemShut {NoStop}%
\bibitem [{\citenamefont {D'Alessio}\ and\ \citenamefont {Rigol}(2014)}]{dalessio14}%
  \BibitemOpen
  \bibfield  {author} {\bibinfo {author} {\bibfnamefont {L.}~\bibnamefont {D'Alessio}}\ and\ \bibinfo {author} {\bibfnamefont {M.}~\bibnamefont {Rigol}},\ }\bibfield  {title} {\bibinfo {title} {{Long-time Behavior of Isolated Periodically Driven Interacting Lattice Systems}},\ }\href {https://doi.org/10.1103/PhysRevX.4.041048} {\bibfield  {journal} {\bibinfo  {journal} {Phys. Rev. X}\ }\textbf {\bibinfo {volume} {4}},\ \bibinfo {pages} {041048} (\bibinfo {year} {2014})}\BibitemShut {NoStop}%
\bibitem [{\citenamefont {Mumford}(2023)}]{mumford23}%
  \BibitemOpen
  \bibfield  {author} {\bibinfo {author} {\bibfnamefont {J.}~\bibnamefont {Mumford}},\ }\bibfield  {title} {\bibinfo {title} {{Many topological regions on the Bloch sphere of the spin-1/2 double-kicked top}},\ }\href {https://doi.org/10.1103/PhysRevA.107.053316} {\bibfield  {journal} {\bibinfo  {journal} {Phys. Rev. A}\ }\textbf {\bibinfo {volume} {107}},\ \bibinfo {pages} {053316} (\bibinfo {year} {2023})}\BibitemShut {NoStop}%
\bibitem [{\citenamefont {Dobrzyniecki}\ and\ \citenamefont {Tomza}(2023)}]{dobryniecki23}%
  \BibitemOpen
  \bibfield  {author} {\bibinfo {author} {\bibfnamefont {J.}~\bibnamefont {Dobrzyniecki}}\ and\ \bibinfo {author} {\bibfnamefont {M.}~\bibnamefont {Tomza}},\ }\bibfield  {title} {\bibinfo {title} {{Quantum simulation of the central spin model with a Rydberg atom and polar molecules in optical tweezers}},\ }\href {https://doi.org/10.1103/PhysRevA.108.052618} {\bibfield  {journal} {\bibinfo  {journal} {Phys. Rev. A}\ }\textbf {\bibinfo {volume} {108}},\ \bibinfo {pages} {052618} (\bibinfo {year} {2023})}\BibitemShut {NoStop}%
\bibitem [{\citenamefont {Ashida}\ \emph {et~al.}(2019)\citenamefont {Ashida}, \citenamefont {Shi}, \citenamefont {Schmidt}, \citenamefont {Sadeghpour}, \citenamefont {Cirac},\ and\ \citenamefont {Demler}}]{demler2019}%
  \BibitemOpen
  \bibfield  {author} {\bibinfo {author} {\bibfnamefont {Y.}~\bibnamefont {Ashida}}, \bibinfo {author} {\bibfnamefont {T.}~\bibnamefont {Shi}}, \bibinfo {author} {\bibfnamefont {R.}~\bibnamefont {Schmidt}}, \bibinfo {author} {\bibfnamefont {H.~R.}\ \bibnamefont {Sadeghpour}}, \bibinfo {author} {\bibfnamefont {J.~I.}\ \bibnamefont {Cirac}},\ and\ \bibinfo {author} {\bibfnamefont {E.}~\bibnamefont {Demler}},\ }\bibfield  {title} {\bibinfo {title} {{Quantum Rydberg Central Spin Model}},\ }\href {https://doi.org/10.1103/PhysRevLett.123.183001} {\bibfield  {journal} {\bibinfo  {journal} {Phys. Rev. Lett.}\ }\textbf {\bibinfo {volume} {123}},\ \bibinfo {pages} {183001} (\bibinfo {year} {2019})}\BibitemShut {NoStop}%
\bibitem [{\citenamefont {Childress}\ \emph {et~al.}(2006)\citenamefont {Childress}, \citenamefont {Dutt}, \citenamefont {Taylor}, \citenamefont {Zibrov}, \citenamefont {Jelezko}, \citenamefont {Wrachtrup}, \citenamefont {Hemmer},\ and\ \citenamefont {Lukin}}]{chldress06}%
  \BibitemOpen
  \bibfield  {author} {\bibinfo {author} {\bibfnamefont {L.}~\bibnamefont {Childress}}, \bibinfo {author} {\bibfnamefont {M.~V.~G.}\ \bibnamefont {Dutt}}, \bibinfo {author} {\bibfnamefont {J.~M.}\ \bibnamefont {Taylor}}, \bibinfo {author} {\bibfnamefont {A.~S.}\ \bibnamefont {Zibrov}}, \bibinfo {author} {\bibfnamefont {F.}~\bibnamefont {Jelezko}}, \bibinfo {author} {\bibfnamefont {J.}~\bibnamefont {Wrachtrup}}, \bibinfo {author} {\bibfnamefont {P.~R.}\ \bibnamefont {Hemmer}},\ and\ \bibinfo {author} {\bibfnamefont {M.~D.}\ \bibnamefont {Lukin}},\ }\bibfield  {title} {\bibinfo {title} {{Coherent Dynamics of Coupled Electron and Nuclear Spin Qubits in Diamond}},\ }\href {https://doi.org/10.1126/science.1131871} {\bibfield  {journal} {\bibinfo  {journal} {Science}\ }\textbf {\bibinfo {volume} {314}},\ \bibinfo {pages} {281} (\bibinfo {year} {2006})}\BibitemShut {NoStop}%
\bibitem [{\citenamefont {Kessler}\ \emph {et~al.}(2010)\citenamefont {Kessler}, \citenamefont {Yelin}, \citenamefont {Lukin}, \citenamefont {Cirac},\ and\ \citenamefont {Giedke}}]{kessler10}%
  \BibitemOpen
  \bibfield  {author} {\bibinfo {author} {\bibfnamefont {E.~M.}\ \bibnamefont {Kessler}}, \bibinfo {author} {\bibfnamefont {S.}~\bibnamefont {Yelin}}, \bibinfo {author} {\bibfnamefont {M.~D.}\ \bibnamefont {Lukin}}, \bibinfo {author} {\bibfnamefont {J.~I.}\ \bibnamefont {Cirac}},\ and\ \bibinfo {author} {\bibfnamefont {G.}~\bibnamefont {Giedke}},\ }\bibfield  {title} {\bibinfo {title} {{Optical Superradiance from Nuclear Spin Environment of Single-Photon Emitters}},\ }\href {https://doi.org/10.1103/PhysRevLett.104.143601} {\bibfield  {journal} {\bibinfo  {journal} {Phys. Rev. Lett.}\ }\textbf {\bibinfo {volume} {104}},\ \bibinfo {pages} {143601} (\bibinfo {year} {2010})}\BibitemShut {NoStop}%
\bibitem [{\citenamefont {Khaetskii}\ \emph {et~al.}(2002)\citenamefont {Khaetskii}, \citenamefont {Loss},\ and\ \citenamefont {Glazman}}]{khaetskii02}%
  \BibitemOpen
  \bibfield  {author} {\bibinfo {author} {\bibfnamefont {A.~V.}\ \bibnamefont {Khaetskii}}, \bibinfo {author} {\bibfnamefont {D.}~\bibnamefont {Loss}},\ and\ \bibinfo {author} {\bibfnamefont {L.}~\bibnamefont {Glazman}},\ }\bibfield  {title} {\bibinfo {title} {{Electron Spin Decoherence in Quantum Dots due to Interaction with Nuclei}},\ }\href {https://doi.org/10.1103/PhysRevLett.88.186802} {\bibfield  {journal} {\bibinfo  {journal} {Phys. Rev. Lett.}\ }\textbf {\bibinfo {volume} {88}},\ \bibinfo {pages} {186802} (\bibinfo {year} {2002})}\BibitemShut {NoStop}%
\bibitem [{\citenamefont {Khaetskii}\ \emph {et~al.}(2003)\citenamefont {Khaetskii}, \citenamefont {Loss},\ and\ \citenamefont {Glazman}}]{khaetskii03}%
  \BibitemOpen
  \bibfield  {author} {\bibinfo {author} {\bibfnamefont {A.}~\bibnamefont {Khaetskii}}, \bibinfo {author} {\bibfnamefont {D.}~\bibnamefont {Loss}},\ and\ \bibinfo {author} {\bibfnamefont {L.}~\bibnamefont {Glazman}},\ }\bibfield  {title} {\bibinfo {title} {{Electron spin evolution induced by interaction with nuclei in a quantum dot}},\ }\href {https://doi.org/10.1103/PhysRevB.67.195329} {\bibfield  {journal} {\bibinfo  {journal} {Phys. Rev. B}\ }\textbf {\bibinfo {volume} {67}},\ \bibinfo {pages} {195329} (\bibinfo {year} {2003})}\BibitemShut {NoStop}%
\bibitem [{bor(2007)}]{bortz07}%
  \BibitemOpen
  \bibfield  {title} {\bibinfo {title} {{Exact dynamics in the inhomogeneous central-spin model}, author = {Bortz, Michael and Stolze, Joachim}},\ }\href {https://doi.org/10.1103/PhysRevB.76.014304} {\bibfield  {journal} {\bibinfo  {journal} {Phys. Rev. B}\ }\textbf {\bibinfo {volume} {76}},\ \bibinfo {pages} {014304} (\bibinfo {year} {2007})}\BibitemShut {NoStop}%
\bibitem [{\citenamefont {Bortz}\ \emph {et~al.}(2010)\citenamefont {Bortz}, \citenamefont {Eggert}, \citenamefont {Schneider}, \citenamefont {St\"ubner},\ and\ \citenamefont {Stolze}}]{bortz10}%
  \BibitemOpen
  \bibfield  {author} {\bibinfo {author} {\bibfnamefont {M.}~\bibnamefont {Bortz}}, \bibinfo {author} {\bibfnamefont {S.}~\bibnamefont {Eggert}}, \bibinfo {author} {\bibfnamefont {C.}~\bibnamefont {Schneider}}, \bibinfo {author} {\bibfnamefont {R.}~\bibnamefont {St\"ubner}},\ and\ \bibinfo {author} {\bibfnamefont {J.}~\bibnamefont {Stolze}},\ }\bibfield  {title} {\bibinfo {title} {{Dynamics and decoherence in the central spin model using exact methods}},\ }\href {https://doi.org/10.1103/PhysRevB.82.161308} {\bibfield  {journal} {\bibinfo  {journal} {Phys. Rev. B}\ }\textbf {\bibinfo {volume} {82}},\ \bibinfo {pages} {161308} (\bibinfo {year} {2010})}\BibitemShut {NoStop}%
\bibitem [{\citenamefont {Shirley}(1965)}]{shirley65}%
  \BibitemOpen
  \bibfield  {author} {\bibinfo {author} {\bibfnamefont {J.~H.}\ \bibnamefont {Shirley}},\ }\bibfield  {title} {\bibinfo {title} {{Solution of the Schr\"odinger Equation with a Hamiltonian Periodic in Time}},\ }\href {https://doi.org/10.1103/PhysRev.138.B979} {\bibfield  {journal} {\bibinfo  {journal} {Phys. Rev.}\ }\textbf {\bibinfo {volume} {138}},\ \bibinfo {pages} {B979} (\bibinfo {year} {1965})}\BibitemShut {NoStop}%
\bibitem [{\citenamefont {Sambe}(1973)}]{sambe73}%
  \BibitemOpen
  \bibfield  {author} {\bibinfo {author} {\bibfnamefont {H.}~\bibnamefont {Sambe}},\ }\bibfield  {title} {\bibinfo {title} {{Steady States and Quasienergies of a Quantum-Mechanical System in an Oscillating Field}},\ }\href {https://doi.org/10.1103/PhysRevA.7.2203} {\bibfield  {journal} {\bibinfo  {journal} {Phys. Rev. A}\ }\textbf {\bibinfo {volume} {7}},\ \bibinfo {pages} {2203} (\bibinfo {year} {1973})}\BibitemShut {NoStop}%
\bibitem [{\citenamefont {Grifoni}\ and\ \citenamefont {Hänggi}(1998)}]{grifoni98}%
  \BibitemOpen
  \bibfield  {author} {\bibinfo {author} {\bibfnamefont {M.}~\bibnamefont {Grifoni}}\ and\ \bibinfo {author} {\bibfnamefont {P.}~\bibnamefont {Hänggi}},\ }\bibfield  {title} {\bibinfo {title} {{Driven quantum tunneling}},\ }\href {https://doi.org/https://doi.org/10.1016/S0370-1573(98)00022-2} {\bibfield  {journal} {\bibinfo  {journal} {Physics Reports}\ }\textbf {\bibinfo {volume} {304}},\ \bibinfo {pages} {229} (\bibinfo {year} {1998})}\BibitemShut {NoStop}%
\bibitem [{\citenamefont {Asb\'{o}th}(2012)}]{asboth12}%
  \BibitemOpen
  \bibfield  {author} {\bibinfo {author} {\bibfnamefont {J.~K.}\ \bibnamefont {Asb\'{o}th}},\ }\bibfield  {title} {\bibinfo {title} {{Symmetries, topological phases, and bound states in the one-dimensional quantum walk}},\ }\href {https://doi.org/10.1103/PhysRevB.86.195414} {\bibfield  {journal} {\bibinfo  {journal} {Phys. Rev. B}\ }\textbf {\bibinfo {volume} {86}},\ \bibinfo {pages} {195414} (\bibinfo {year} {2012})}\BibitemShut {NoStop}%
\bibitem [{\citenamefont {Asb\'{o}th}\ and\ \citenamefont {Hideaki}(2013)}]{asboth13}%
  \BibitemOpen
  \bibfield  {author} {\bibinfo {author} {\bibfnamefont {J.~K.}\ \bibnamefont {Asb\'{o}th}}\ and\ \bibinfo {author} {\bibfnamefont {O.}~\bibnamefont {Hideaki}},\ }\bibfield  {title} {\bibinfo {title} {{Bulk-boundary correspondence for chiral symmetric quantum walks}},\ }\href {https://doi.org/10.1103/PhysRevB.88.121406} {\bibfield  {journal} {\bibinfo  {journal} {Phys. Rev. B}\ }\textbf {\bibinfo {volume} {88}},\ \bibinfo {pages} {121406(R)} (\bibinfo {year} {2013})}\BibitemShut {NoStop}%
\bibitem [{\citenamefont {Zirnbauer}(1996)}]{zirnbauer96}%
  \BibitemOpen
  \bibfield  {author} {\bibinfo {author} {\bibfnamefont {M.~R.}\ \bibnamefont {Zirnbauer}},\ }\bibfield  {title} {\bibinfo {title} {{Riemannian symmetric superspaces and their origin in random‐matrix theory}},\ }\href {https://doi.org/10.1063/1.531675} {\bibfield  {journal} {\bibinfo  {journal} {J. Math. Phys.}\ }\textbf {\bibinfo {volume} {37}},\ \bibinfo {pages} {4986} (\bibinfo {year} {1996})}\BibitemShut {NoStop}%
\bibitem [{\citenamefont {Altland}\ and\ \citenamefont {Zirnbauer}(1997)}]{altland97}%
  \BibitemOpen
  \bibfield  {author} {\bibinfo {author} {\bibfnamefont {A.}~\bibnamefont {Altland}}\ and\ \bibinfo {author} {\bibfnamefont {M.~R.}\ \bibnamefont {Zirnbauer}},\ }\bibfield  {title} {\bibinfo {title} {{Nonstandard symmetry classes in mesoscopic normal-superconducting hybrid structures}},\ }\href {https://doi.org/10.1103/PhysRevB.55.1142} {\bibfield  {journal} {\bibinfo  {journal} {Phys. Rev. B}\ }\textbf {\bibinfo {volume} {55}},\ \bibinfo {pages} {1142} (\bibinfo {year} {1997})}\BibitemShut {NoStop}%
\bibitem [{\citenamefont {Heinzner}\ \emph {et~al.}(2005)\citenamefont {Heinzner}, \citenamefont {Huckleberry},\ and\ \citenamefont {Zirnbauer}}]{heinzner05}%
  \BibitemOpen
  \bibfield  {author} {\bibinfo {author} {\bibfnamefont {P.}~\bibnamefont {Heinzner}}, \bibinfo {author} {\bibfnamefont {A.}~\bibnamefont {Huckleberry}},\ and\ \bibinfo {author} {\bibfnamefont {M.~R.}\ \bibnamefont {Zirnbauer}},\ }\bibfield  {title} {\bibinfo {title} {{Symmetry Classes of Disordered Fermions}},\ }\href {https://doi.org/10.1007/s00220-005-1330-9} {\bibfield  {journal} {\bibinfo  {journal} {Commun. Math. Phys.}\ }\textbf {\bibinfo {volume} {257}},\ \bibinfo {pages} {725} (\bibinfo {year} {2005})}\BibitemShut {NoStop}%
\bibitem [{\citenamefont {Song}\ and\ \citenamefont {Prodan}(2014)}]{song14}%
  \BibitemOpen
  \bibfield  {author} {\bibinfo {author} {\bibfnamefont {J.}~\bibnamefont {Song}}\ and\ \bibinfo {author} {\bibfnamefont {E.}~\bibnamefont {Prodan}},\ }\bibfield  {title} {\bibinfo {title} {{AIII and BDI topological systems at strong disorder}},\ }\href {https://doi.org/10.1103/PhysRevB.89.224203} {\bibfield  {journal} {\bibinfo  {journal} {Phys. Rev. B}\ }\textbf {\bibinfo {volume} {89}},\ \bibinfo {pages} {224203} (\bibinfo {year} {2014})}\BibitemShut {NoStop}%
\bibitem [{\citenamefont {Mondragon-Shem}\ \emph {et~al.}(2014)\citenamefont {Mondragon-Shem}, \citenamefont {Hughes}, \citenamefont {Song},\ and\ \citenamefont {Prodan}}]{shem14}%
  \BibitemOpen
  \bibfield  {author} {\bibinfo {author} {\bibfnamefont {I.}~\bibnamefont {Mondragon-Shem}}, \bibinfo {author} {\bibfnamefont {T.~L.}\ \bibnamefont {Hughes}}, \bibinfo {author} {\bibfnamefont {J.}~\bibnamefont {Song}},\ and\ \bibinfo {author} {\bibfnamefont {E.}~\bibnamefont {Prodan}},\ }\bibfield  {title} {\bibinfo {title} {{Topological criticality in the chiral symmetric AIII class at strong disorder}},\ }\href {https://doi.org/10.1103/PhysRevLett.113.046802} {\bibfield  {journal} {\bibinfo  {journal} {Phys. Rev. Lett.}\ }\textbf {\bibinfo {volume} {113}},\ \bibinfo {pages} {046802} (\bibinfo {year} {2014})}\BibitemShut {NoStop}%
\bibitem [{\citenamefont {Sieberer}\ \emph {et~al.}(2019)\citenamefont {Sieberer}, \citenamefont {Olsacher}, \citenamefont {Elben}, \citenamefont {Heyl}, \citenamefont {Hauke}, \citenamefont {Haake},\ and\ \citenamefont {Zoller}}]{sieberer19}%
  \BibitemOpen
  \bibfield  {author} {\bibinfo {author} {\bibfnamefont {L.~M.}\ \bibnamefont {Sieberer}}, \bibinfo {author} {\bibfnamefont {T.}~\bibnamefont {Olsacher}}, \bibinfo {author} {\bibfnamefont {A.}~\bibnamefont {Elben}}, \bibinfo {author} {\bibfnamefont {M.}~\bibnamefont {Heyl}}, \bibinfo {author} {\bibfnamefont {P.}~\bibnamefont {Hauke}}, \bibinfo {author} {\bibfnamefont {F.}~\bibnamefont {Haake}},\ and\ \bibinfo {author} {\bibfnamefont {P.}~\bibnamefont {Zoller}},\ }\bibfield  {title} {\bibinfo {title} {{Digital quantum simulation, Trotter errors, and quantum chaos of the kicked top}},\ }\href {https://doi.org/10.1038/s41534-019-0192-5} {\bibfield  {journal} {\bibinfo  {journal} {npj Quauntum Information}\ }\textbf {\bibinfo {volume} {5}},\ \bibinfo {pages} {78} (\bibinfo {year} {2019})}\BibitemShut {NoStop}%
\bibitem [{\citenamefont {Sen}\ \emph {et~al.}(2021)\citenamefont {Sen}, \citenamefont {Sen},\ and\ \citenamefont {Sengupta}}]{sen21}%
  \BibitemOpen
  \bibfield  {author} {\bibinfo {author} {\bibfnamefont {A.}~\bibnamefont {Sen}}, \bibinfo {author} {\bibfnamefont {D.}~\bibnamefont {Sen}},\ and\ \bibinfo {author} {\bibfnamefont {K.}~\bibnamefont {Sengupta}},\ }\bibfield  {title} {\bibinfo {title} {{Analytic approaches to periodically driven closed quantum systems: methods and applications}},\ }\href {https://doi.org/10.1088/1361-648X/ac1b61} {\bibfield  {journal} {\bibinfo  {journal} {J. Phys. A: Math. Gen.}\ }\textbf {\bibinfo {volume} {33}},\ \bibinfo {pages} {443003} (\bibinfo {year} {2021})}\BibitemShut {NoStop}%
\bibitem [{\citenamefont {Deutsch}(1991)}]{deutsch91}%
  \BibitemOpen
  \bibfield  {author} {\bibinfo {author} {\bibfnamefont {J.~M.}\ \bibnamefont {Deutsch}},\ }\bibfield  {title} {\bibinfo {title} {{Quantum Statistical Mechanics in a Closed System}},\ }\href {https://doi.org/10.1103/PhysRevA.43.2046} {\bibfield  {journal} {\bibinfo  {journal} {Phys. Rev. A}\ }\textbf {\bibinfo {volume} {43}},\ \bibinfo {pages} {2046} (\bibinfo {year} {1991})}\BibitemShut {NoStop}%
\bibitem [{\citenamefont {Srednicki}(1994)}]{srednicki94}%
  \BibitemOpen
  \bibfield  {author} {\bibinfo {author} {\bibfnamefont {M.}~\bibnamefont {Srednicki}},\ }\bibfield  {title} {\bibinfo {title} {{Chaos and Quantum Thermalization}},\ }\href {https://doi.org/10.1103/PhysRevE.50.888} {\bibfield  {journal} {\bibinfo  {journal} {Phys. Rev. E}\ }\textbf {\bibinfo {volume} {50}},\ \bibinfo {pages} {888} (\bibinfo {year} {1994})}\BibitemShut {NoStop}%
\bibitem [{\citenamefont {Srednicki}(1999)}]{srednicki99}%
  \BibitemOpen
  \bibfield  {author} {\bibinfo {author} {\bibfnamefont {M.}~\bibnamefont {Srednicki}},\ }\bibfield  {title} {\bibinfo {title} {{Chaos and Quantum Thermalization}},\ }\href {https://doi.org/10.1088/0305-4470/32/7/007} {\bibfield  {journal} {\bibinfo  {journal} {J. Phys. A: Math. Gen.}\ }\textbf {\bibinfo {volume} {32}},\ \bibinfo {pages} {1163} (\bibinfo {year} {1999})}\BibitemShut {NoStop}%
\bibitem [{Note1()}]{Note1}%
  \BibitemOpen
  \bibinfo {note} {This condition rigorously justifies the rotating-wave approximation but when studying bound states away from the equatorial plane of the top's Bloch sphere it may be sufficient to relax it to $\Omega \gg \Delta , \protect \sqrt {2j}\chi $.}\BibitemShut {Stop}%
\bibitem [{\citenamefont {Gamel}\ and\ \citenamefont {James}(2010)}]{gamel10}%
  \BibitemOpen
  \bibfield  {author} {\bibinfo {author} {\bibfnamefont {O.}~\bibnamefont {Gamel}}\ and\ \bibinfo {author} {\bibfnamefont {D.~F.~V.}\ \bibnamefont {James}},\ }\bibfield  {title} {\bibinfo {title} {{Time-averaged quantum dynamics and the validity of the effective Hamiltonian model}},\ }\href {https://doi.org/10.1103/PhysRevA.82.052106} {\bibfield  {journal} {\bibinfo  {journal} {Phys. Rev. A}\ }\textbf {\bibinfo {volume} {82}},\ \bibinfo {pages} {052106} (\bibinfo {year} {2010})}\BibitemShut {NoStop}%
\end{thebibliography}%

\end{document}